\RequirePackage[2020-02-02]{latexrelease}
\documentclass[superscriptaddress,secnumarabic,
amssymb,amsmath,nobibnotes,aps,prd,showkeys,showpacs,nofootinbib]{revtex4}%
\usepackage{graphicx}
\usepackage{tabularx}
\usepackage{epsf}
\usepackage{epsfig}
\usepackage{scrtime}
\usepackage{adjustbox}
\usepackage[multiple]{footmisc}
\usepackage{amsfonts, graphics}
\usepackage{bm}
\usepackage[utf8x]{inputenc}
\usepackage{amsmath}
\usepackage{amsfonts}
\usepackage{amssymb}
\usepackage{color}
\usepackage{epstopdf}
\usepackage{natbib}%
\usepackage{multirow}
\usepackage{array}
\usepackage{hhline}
\usepackage{wrapfig}
\usepackage{lipsum}
\usepackage{showframe}
\newcolumntype{C}{>{\centering\arraybackslash}X}
\providecommand{\U}[1]{\protect\rule{.1in}{.1in}}

\begin{document}

\title{Some Classes of Interacting Two-Fluid Model of the Expanding Universe}
\author{Subhra Bhattacharya}
\email{subhra.maths@presiuniv.ac.in}
\affiliation{Department of Mathematics, Presidency University, Kolkata-700073, India}


\begin{abstract}

We consider interacting dark matter-dark energy models arising out of a general interaction term $Q=f(\rho_{m},\rho_{d},\dot{\rho}_{m},\dot{\rho}_{d}).$ Here $f$ is a functional relation connecting the energy densities $\rho_{m}$ and $\rho_{d}$ and their derivatives w.r.t. time $t.$ In our model we consider two interacting barotropic fluid with constant equation of state $\omega_{m}$ and $\omega_{d}.$ By considering a dynamical interaction between them we trace out the cosmological evolution dynamics of the universe. We analytically solve the model by considering a constant ratio between the two fluids and then track the corresponding analytical results using observational data from the baryon acoustic oscillation measurements, Type Ia supernovae measurements and the local Hubble constant measurements. From this general setting we introduce three different models and nine different interaction function. Our final aim is to set up a comparative analysis of the various class of models under the different interaction function using common theoretical and numerical analysis. 
\end{abstract}

\vspace{1em}

\maketitle

\vspace{1em}

\section{Introduction}

Observational data from distant supernovae and cosmic microwave background have established that the universe is undergoing accelerated expansion \cite{sup1,sup2}. Theoretical predictions suggest that the acceleration is driven by some matter component with a negative pressure that is sufficient to counter the attractive gravitational force. Considering a homogeneous and isotropic universe, studies show that the primary matter component of our universe is the dark energy (DE), characterized by negative pressure while most of the remaining matter is made up of pressure-less dark matter (DM). The rest of the universe comprises of  negligible amount of baryonic matter and radiations. Such a universe is modelled by the simple concordance model or the $\Lambda$CDM model \cite{lcdm}, that best fits the observational data. Infact $\Lambda$CDM is so successful that a locally inhomogeneous universe can mimic $\Lambda$CDM cosmology at large scales \cite{sb1}. In the $\Lambda$CDM model the expansion of the universe is attributed to cosmological constant $\Lambda$ with negative pressure \cite{lam}. $\Lambda$ corresponds to the vacuum energy density, resulting from vacuum fluctuations of matter with equation of state $\omega_{\Lambda}=\frac{p_{\Lambda}}{\rho_{\Lambda}}=-1.$ 

Despite the phenomenal success of the $\Lambda$CDM model, it is mired with several theoretical limitations. Two important, yet unresolved puzzles are the cosmological constant (CC) problem \cite{cc} and the coincidence problem \cite{coin}. Calculation reveal a 121 order of magnitude discrepancy in the numerical estimates of the cosmological constant when compared with the actual estimate of vacuum energy density. This difference between the numerical and theoretical estimates is called the cosmological constant problem. The second perplexing puzzle is the similar orders of magnitude of the DE and DM energy density in the current epoch. The densities of different matter components in the universe did not remain same throughout its evolution. The hot big bang divides the evolution into three phases. An initial hot radiation dominated epoch. This was followed by the DM dominated epoch, during which most of the structures we observe in the universe were created. Curiously this changed at a recent time around red-shift distance of $z\approx 0.5$ when densities of DM and DE became equivalent. During the first two phases of evolution the density of DE component was negligible in the universe, yet it is it only by coincidence that in the current epoch it evolved to similar orders of magnitude as the other matter components. This is the coincidence problem. 

A considerable bulk of the literature has been devoted to tackle these inconsistencies, yet the questions remain. Models like scalar field description of DE, modified gravity theories like $f(R), f(T)$ and recently $f(Q)$ \cite{mgt}, the dynamical DE models with variable equation of state parameter (also called $\omega CDM$), cosmic fluids etc. have been used time and again to resolve these issues \cite{rev}. One such theoretical model is the interacting DE and DM description of the universe \cite{ide1, ide2, ide3, ide4,pwan, am, guo, he, mvg, liu, 1c2, 2c2, 3c2, 4c2, c51, c52, c53, sb2, c81, sb3, c9, sp, wyan, iw}. Most cosmological models assume independent evolution for DE and DM. It was shown that scalar field models can couple with other matter fields via energy transfer between the various matter sectors. Various work showed that models depicting viable interaction between DE and DM could help to resolve the cosmological constant problem \cite{idcc}. Since the current energy budget of the universe show predominance of the two dark components, it is not unusual to assume non-gravitational interaction between them resulting in their simultaneous evolution, which in turn can lead to successful resolution of the coincidence problem \cite{idcoin}. Several models of interacting DE and DM has been suggested in the literature. For a complete review see \cite{idrev}. Recent studies have also predicted that interacting models can have significant contribution in easing the present tension in Hubble constant $H_{0}$ \cite{ht, 1c2, 2c2}. Due to the variety of possibilities, the interacting models are some of the extensively studied models in the literature \cite{rec2}-\cite{rec36}. Most models predict minimal phenomenological interaction, involving the Hubble parameter, the fluid energy densities and sometimes the derivatives of the dark fluid energy densities.  

In this article we shall consider one such general interaction function $Q=f(\rho_{m},\rho_{d},\dot{\rho}_{m},\dot{\rho}_{d},H).$ Here $\rho_{m}$ and $\rho_{d}$ denote the two interacting dark fluids and $\dot{\rho}_{m},\dot{\rho}_{d}$ denote their derivatives w.r.t time $t.$ These two dark fluid interact via non gravitational coupling term $Q$. In the limiting case of $Q\rightarrow 0$ we can get back the usual non-interacting scenario. It may be noted that the choice of interaction function is based purely on phenomenological considerations. Attempts to obtain suitable interaction based on underlying physics has so far produced no successful results. Since interacting DM-DE cosmologies track the background dynamics of the universe, it is expected that interactions will have dependence on the Hubble parameter and and the two dark densities affecting the cosmology. Hence it is logical to assume a functional relation between them and their rate of evolution. We therefore choose a general interaction having dependence on all the factors that can affect interaction, namely $H,\rho_{m},\rho_{d}$ and their rate of change. We solve the ensuing system by theoretical means with the assumption of a constant ratio between DE and DM. We divide the resulting solution into nine subclasses of solution, depending upon specific parameter choices. These nine sub-classes are constrained using binned Supernovae data from JLA light curves, the Hubble and BAO data in three separate models. The various parameter constraints are used to set up a comparative analysis between the several interaction functions. We also trace the evolution of the equation of state the effective matter and the corresponding dynamical deceleration parameter. Finally we do statistical fitness analysis among the various classes of interaction function considered to understand which models to accept and which ones to rule out.

The article is arranged as follows: in section 2 we provide details of the relevant equations, the analytical results, the interaction functions and the model formulations. In section 3 we provide details of the data sets used with the details and discussion of the corresponding results obtained. In section 4 we discuss the effective equation of state, while in section 5 the deceleration parameter for the model. In section 6 we provide statistical model selection details and finally in section 7 we end with a brief discussion.

\section{Modelling the Interacting Dark Sector}

We consider a universe described by the flat Friedmann–Lemaître–Robertson–Walker metric (FLRW) given by:
\begin{equation}
dS^{2}=dt^{2}-a^{2}(t)\left(dx^{2}+dy^{2}+dz^{2}\right)
\end{equation}
where $t$ is the cosmic time and $a(t)$ is the scale factor. The stress energy tensors consist of coupled barotropic fluids $\rho_{m}$ and $\rho_{d}$ with matter distribution $T_{\mu\nu}$ following the general equation:
\begin{equation}
T_{\mu\nu}=(p+\rho)u_{\mu}u_{\nu}+pg_{\mu\nu}
\end{equation}
with $u_{\mu}$ the four velocity of the fluid and $p$ being the fluid pressure. The Einstein's field equations are given by:
\begin{align}
3H^{2}&=8\pi G(\rho_{m}+\rho_{d})\label{f1}\\
-2\dot{H}&=8\pi G(\rho_{m}+p_{m}+\rho_{d}+p_{d})\label{f2}
\end{align}
where with the Hubble parameter $H=\frac{\dot{a}}{a}$ and \emph{dot} denoting derivative w.r.t time $t$. We consider that the interacting dark sectors have equation of state $p_{m}=\omega_{m}\rho_{m}$ and $p_{d}=\omega_{d}\rho_{d}$ with both $\omega_{m}$ and $\omega_{d}$ being constants. Since the total energy density is conserved, the continuity equations evolve as: 
\begin{align}
\dot{\rho_{m}}+3H\rho_{m}(1+\omega_{m})=Q\\
\dot{\rho_{d}}+3H\rho_{d}(1+\omega_{d})=-Q
\end{align} 
with $Q$ being the interaction function. Here a $Q>0$ term indicate an energy transfer from DE to DM, while $Q<0$ indicates transfer of energy from DM to DE. Let $\rho_{t}=\rho_{m}+\rho_{d}$ be the total energy density of the combined dark sectors. Then from the above continuity equations we observe that $\rho_{t}$ follows the conservation equation $\dot{\rho_{t}}+3H\rho_{t}(1+\omega_{t})=0$ with $\omega_{t}=\frac{\omega_{m}\rho_{m}+\rho_{d}\omega_{d}}{\rho_{m}+\rho_{d}}.$
The continuity equations are modified using $y=3\log a$ to give
\begin{align}
\rho_{m}'+\rho_{m}(1+\omega_{m})=R\label{c1}\\
\rho_{d}'+\rho_{d}(1+\omega_{d})=-R\label{c2}
\end{align}
with dash denoting differentiation w.r.t the new variable $y$ and $R=\frac{Q}{3H}.$ Here we consider the interaction as a linear combination of the factors affecting the background dynamics and is given by:
\begin{equation}
R=\xi\rho_{m}+\eta\rho_{d}+\alpha\rho'_{m}+\beta\rho'_{d}+\gamma\rho^{\delta}_{m}\rho^{1-\delta}_{d}\label{int}
\end{equation}
where parameters $\xi,~\eta,~\alpha,~\beta,~\gamma$ and $\delta$ are free constants that will be determined using observational data. With 
\begin{equation}
\rho_{d}=r\rho_{m}\label{rat}
\end{equation} we solve (\ref{c1}) and (\ref{c2}) and obtain :
\begin{align}
\rho_{m}=\rho_{m0}(1+z)^{3\left(\frac{1+\omega_{m}-\xi-\eta r-\gamma r^{1-\delta}}{1-\alpha-\beta r}\right)}\label{rm}\\
\rho_{d}=\rho_{d0}(1+z)^{3\left(\frac{1+\omega_{d}+\frac{\xi}{r}+\eta+\frac{\gamma}{r^{\delta}}}{1+\frac{\alpha}{r}+\beta}\right)}\label{rd}
\end{align}
where $1+z=\frac{1}{a}$ is the red-shift distance. Using the Friedmann's equation (\ref{f1}) we obtain that the evolution dynamics of the universe in terms of the Hubble's parameter as:
\begin{equation}
3H^{2}=8\pi G\left(\rho_{m0}(1+z)^{3\left(\frac{1+\omega_{m}-\xi-\eta r-\gamma r^{1-\delta}}{1-\alpha-\beta r}\right)}+\rho_{d0}(1+z)^{3\left(\frac{1+\omega_{d}+\frac{\xi}{r}+\eta+\frac{\gamma}{r^{\delta}}}{1+\frac{\alpha}{r}+\beta}\right)}\right)\label{expan1}
\end{equation}
In terms of the density parameters the above expression can be rewritten as:
\begin{equation}
E^{2}(z)=\Omega_{m}(1+z)^{3\left(\frac{1+\omega_{m}-\xi-\eta r-\gamma r^{1-\delta}}{1-\alpha-\beta r}\right)}+\Omega_{d}(1+z)^{3\left(\frac{1+\omega_{d}+\frac{\xi}{r}+\eta+\frac{\gamma}{r^{\delta}}}{1+\frac{\alpha}{r}+\beta}\right)}\label{expan1}
\end{equation}
where $E(z)=\frac{H(z)}{H_{0}}$ is the dimensionless expansion coefficient ($H_{0}$ is the Hubble parameter at the present time), $\Omega_{m}=\frac{\rho_{m0}}{\rho_{c0}},~\Omega_{d}=\frac{\rho_{d0}}{\rho_{c0}}$ with $\rho_{c0}=\frac{3H_{0}^{2}}{8\pi G}$ and $\Omega_{m}+\Omega_{d}=1.$ We can obtain different interacting functions from (\ref{expan1}) by setting the free parameters to zero in order. Accordingly this general function has been split up into nine different interacting functions as below:

\begin{itemize}
\item {\bf C1: $R=\xi\rho_{m}$} \\
Here $(\eta,~\alpha,~\beta,~\gamma,~\delta)=0$ with the expansion term
$E^{2}(z)=(1-\Omega_{d})(1+z)^{3(1+\omega_{m}-\xi)}+\Omega_{d}(1+z)^{3(1+\omega_{d}+\frac{\xi}{r})}.$

\item {\bf C2: $R=\eta\rho_{d}$} \\
Here we take $(\xi,~\alpha,~\beta,~\gamma,~\delta)=0$ with the expansion factor
$E^{2}(z)=(1-\Omega_{d})(1+z)^{3(1+\omega_{m}-\eta r)}+\Omega_{d}(1+z)^{3(1+\omega_{d}+\eta)}.$ 

\item {\bf C3: $R=\xi\rho_{m}+\alpha\rho'_{m}$} \\
Here we take $(\eta,~\beta,~\gamma,~\delta)=0.$ This gives the expansion coefficient as:
$E^{2}(z)=(1-\Omega_{d})(1+z)^{3\left(\frac{1+\omega_{m}-\xi}{1-\alpha}\right)}+\Omega_{d}(1+z)^{3\left(\frac{1+\omega_{d}+\xi/r}{1+\alpha/r}\right)}.$ 

\item {\bf C4: $R=\eta\rho_{d}+\beta\rho'_{d}$} \\
In this case we take $(\xi,~\alpha,~\gamma,~\delta)=0$ with
$E^{2}(z)=(1-\Omega_{d})(1+z)^{3\left(\frac{1+\omega_{m}-\eta r}{1-\beta r}\right)}+\Omega_{d}(1+z)^{3\left(\frac{1+\omega_{d}+\eta}{1+\beta}\right)}.$ 

\item {\bf C5: $R=\xi\rho_{m}+\eta\rho_{d}$} \\
Here we take $(\alpha,~\beta,~\gamma,~\delta)=0$ with
$E^{2}(z)=(1-\Omega_{d})(1+z)^{3(1+\omega_{m}-\xi-\eta r)}+\Omega_{d}(1+z)^{3(1+\omega_{d}+\xi/r+\eta)}.$ 

\item {\bf C6: $R=\eta\rho_{d}+\alpha\rho'_{m}$}\\
 For this we assume $(\xi,~\beta,~\gamma,~\delta)=0$ and the expansion given by:
$E^{2}(z)=(1-\Omega_{d})(1+z)^{3\left(\frac{1+\omega_{m}-\eta r}{1-\alpha}\right)}+\Omega_{d}(1+z)^{3\left(\frac{1+\omega_{d}+\eta}{1+\alpha/r}\right)}.$ 

\item {\bf C7: $R=\xi\rho_{m}+\beta\rho'_{d}$}\\
 Here we take $(\eta,~\alpha,~\gamma,~\delta)=0$ and the expansion factor reduces to:
$E^{2}(z)=(1-\Omega_{d})(1+z)^{3\left(\frac{1+\omega_{m}-\xi}{1-\beta r}\right)}+\Omega_{d}(1+z)^{3\left(\frac{1+\omega_{d}+\xi/r}{1+\beta}\right)}.$ 

\item {\bf C8: $R=\alpha\rho'_{m}+\beta\rho'_{d}$}\\
 Here $(\xi,~\eta,~\gamma,~\delta)=0$ and the expansion factor is:
$E^{2}(z)=(1-\Omega_{d})(1+z)^{3\left(\frac{1+\omega_{m}}{1-\alpha-\beta r}\right)}+\Omega_{d}(1+z)^{3\left(\frac{1+\omega_{d}}{1+\alpha/r+\beta}\right)}.$ 

\item {\bf C9: $R=\gamma\rho_{m}^{\delta}\rho_{d}^{1-\delta}$} Finally we take $(\xi,~\eta,~\alpha,~\beta)=0$ which gives:
$E^{2}(z)=(1-\Omega_{d})(1+z)^{3\left(1+\omega_{m}-\gamma r^{1-\delta}\right)}+\Omega_{d}(1+z)^{3\left(1+\omega_{d}+\gamma r^{-\delta}\right)}.$ 

\end{itemize}

We shall consider these nine independent interaction functions within the purview of three different cosmological models. 
\begin{itemize}
\item[{\bf M1:}] \emph{Interacting Cold Dark Matter and Cosmological Constant}\\
We will consider $\rho_{m}$ as the pressure-less cold dark matter (CDM) with $\omega_{m}=0$ and $\rho_{d}$ as the cosmological constant (CC) with $\omega_{d}=-1$. 
\item[{\bf M2:}] \emph{Interacting Cold Dark Matter and Dark Energy} \\
Here although $\rho_{m}$ is considered to be CDM, $\rho_{d}$ is considered to be a fluid with free equation of state parameter $\omega_{d}.$ 
\item[{\bf M3:}] \emph{Interacting Two Fluid} \\  
In this model we consider interaction between two fluid with constant equation of state $\omega_{m}$ and $\omega_{d}$ which are which are free to be determined using the observational data sets.
\end{itemize}

We constrain the free parameters in the nine interaction functions using the models M1-M3. Based on the results we set up a comparison between the different models and the interaction functions. Although one can find numerous literature on coupled DE and DM cosmology, yet we cannot bring them under a single umbrella due to diversity in solution approach and data sets. This work thus aims to bring a fairly large class of coupled DE-DM models using a unified theoretical and numerical approach for their better understanding and classification. Further, analysing these scenarios in three different models will help us underline the most convincing form of DE and DM that supports coupled evolution of the two most prominent energy sources of the universe.

\section{Data Sets Used}

\begin{description}
\item {\it JLA Data}: We used a sample of 31 binned data and their corresponding covariance matrix data from the Joint Light-curve analysis \cite{jla}. The corresponding $\chi^{2}$ function that is used to constrain the theoretical model is given by: 
\begin{equation}
\chi^{2}_{JLA}=(\boldsymbol{\mu^{obs}}-\boldsymbol{\mu})^{T}\boldsymbol{C}_{\mu}^{-1}(\boldsymbol{\mu^{*}}-\boldsymbol{\mu})
\end{equation}
with $\mu(z)$ being the distance modulus of type Ia supernova located at redshift $z$ and $\boldsymbol{C}_{\mu}$ the covariance matrix for the binned JLA data set.

\item {\it OHD Data:} We have used 35 sets of observed Hubble data corresponding to the red shift range $0.07\leq z\leq 2.36.$ The data sets have been derived from Cosmic Chronometers \cite{ccr}, the baryon acoustic oscillation method \cite{al} and the baryon acoustic oscillations in the Ly$\alpha$ forest of high red shift quasars \cite{del}. The corresponding $\chi^{2}$ is defined as:
\begin{equation}
\chi^{2}_{OHD}=\sum_{i=1}^{35}\frac{\left(H^{obs}(z_{i})-H(z_{i}(\Theta))\right)^{2}}{\sigma_{i}^{2}}
\end{equation}

\item {\it BAO Data:} The BAO data used in the analysis were obtained from the following sources: 
\begin{enumerate}
\item 6dF Galaxy Survey at $z=0.106$ \cite{df},
\item the SDSS DR7 main galaxy sample at $z=0.15$  \cite{sd},
\item the BOSS LOWZ data at $z=0.32$ and
\item BOSS CMASS at $z=0.57$  \cite{boss}.
\end{enumerate}
The corresponding $\chi^{2}$ is given by:
\begin{equation}
\chi^{2}_{BAO}=\sum_{i=1}^{4}\frac{(\boldsymbol{(\eta}^{obs}(z_{i})-\boldsymbol{\eta}(z_{i}\Theta))^{2}}{\boldsymbol{\sigma}_{i}^{2}}
\end{equation}
where $\boldsymbol{\eta}(z_{i})$ is related to the ratio of $d_{V}$ the volume dilation scale and $r_{s},$ the co-moving size of sound horizon at baryon drag epoch. The volume dilation scale $d_{V}$ is related to the luminosity distance $d_{L}$ by $d_{V}=\left(\frac{zd_{L}^{2}}{(1+z)^{2}H(z)}\right)^{1/3}.$ 

$\Theta$ in above three expressions indicate the various parameter set constrained in the models.

\end{description}

\subsection{Statistical Method Applied}

We use the Markov Chain Monte Carlo (MCMC) technique of parameter estimation using the $Python$ application of the MCMC that was devised by Goodman and Weare \cite{gd} and Foreman-Mackey {\it et al} \cite{fr}. Using the Bayes' theorem we can relate the posterior probability distribution of the parameter space to its prior probability distribution and determine the corresponding likelihood function. The minimum $\chi^{2}$ or the maximum likelihood function is given by:
\begin{equation}
-\ln \mathfrak{L}(\Theta)=\frac{1}{2}\left(\chi^{2}_{JLA}+\chi^{2}_{OHD}+\chi^{2}_{BAO}\right)
\end{equation}
where $\Theta$ denotes the set of parameters.

\subsection{Model Analysis}

We have used one data set consisting of a combined JLA+OHD+BAO data to constrain the free parameters $r,\xi,\eta, \alpha,\beta, \gamma,\delta,\omega_{d},\omega_{m},\Omega_{d}.$ For model M1 we have the ratio factor $r,$ the cosmological constant energy density $\Omega_{d}$ and the interaction coefficients as parameters, while for M2 we have the DE equation of state $\omega_{d}$ as an added parameter and model M3 we have both DE and DM energy densities $\omega_{d},~\omega_{m}$ as additional free parameters. In Table I-Table IX we give a tabular representation of the constrained parameters, while in Figures 1-27 (provided in Appendix A) we give the corresponding confidence contours on the 2 dimensional parameter space. 

Table I, Fig.1 - Fig.3 shows the data for interaction function C1. For this interaction all models predict marginal interaction between the two matter sectors with $\xi=0.04$ for M1 and M2 and 0.08 for M3. Further in all the models the $\Omega_{d}$ or the DE density matches with Planck Survey 2018 predictions \cite{plnk}. Model M2 is found converging towards cosmological constant DE candidate with $\omega_{d}$ fitted to a value of -1.00. However in M3 where both DM and DE is left free it is seen that the model has tendency to predict a marginally non-cold DM candidate with $\omega_{m}=0.04,$ and DE favouring the phantom regime. The ratio $r$ of DE and DM, which can also be called the coincidence factor when left free always has a tendency to hover around $2.2-2.3.$ The interaction C1 with model M2 has been is one of the best studied interaction in the literature \cite{pwan, am, guo, he, mvg, liu}. Comparing the results of C1 in M2 with results in \cite{guo}, we find our results being completely compatible. The authors in \cite{guo} used WMAP CMB data, SDSS BAO data and 71 Supernova Legacy Survey data. In the combined analysis they found that the interaction varied as $(-0.4,0.3)$ while the DE density $\omega_{d}$ varied between (-1.16,-0.91). More recently in \cite{liu}, the authors study C1 in M2 in a different context using latest, CMB, Lensing, Pantheon and BAO data. Again their combined analysis constrains the interaction to a slightly positive value with $\omega_{d}$ constrained at -1.019. This clearly shows how C1 compares in different models that we have used and with the existing data.   

Table II, Fig.4 - Fig.6 show the corresponding data for interaction function C2. All models predict marginal interaction with M3 favouring marginally more interaction than M1 and M2 and the DE density $\Omega_{d}$ converge towards Planck 2018 predicted value. Both M2 and M3 with C2 favours marginal phantom DE while from M3 we see that marginal non-cold DM is favoured. The coincidence factor for M1-M3 closely resembles, with $r$ value fitted same for M2 and M3. Previously C2 has been studied in \cite{he, mvg}. Recently one can find renewed interest in the interaction from the bulk of recent literature  \cite{1c2,2c2,3c2,4c2}. In \cite{4c2} the authors use C2 in both M1 and M2. They constrain their results using latest CMB, BAO and Pantheon data. In the combined analysis for M1 they obtain a mean positive interaction value of 0.032, while in M2, they use two different models one for $\omega_{d}<-1$ and other for $\omega_{d}>-1.$ For the $\omega_{d}<-1$ case they obtain a slightly negative interaction with constrained mean $\omega_{d}=-1.08.$ Again we see excellent comparison with our results. 

Table III, Fig.7 - Fig.9 shows the constrained parameter data for function C3. Here C3 is driven via $\rho_{m}$ and $\rho'_{m}.$ In all models the DE density parameter $\Omega_{d}$ is fitted to values that closely mimics recent Planck results. In both M1 and M2 we see DE converges to cosmological constant, while the coincidence factor $r\simeq 2.3.$ The interaction parameters all converge towards negative values with interaction coefficients no longer marginal.  

The results for C4 is shown in table IV and Fig.10 - Fig.12. Here interaction is driven via $\rho_{d}$ and $\rho'_{d}.$ Clearly this interaction does not favour CDM and CC as coupled matter components as is evident from the results in M1. Further results of M2 and M3 clearly shows preference to a quintessence DE candidate. In M3 we again see that C4 also favouring a marginal non-cold DM candidate. $r$ values in M2 and M3 are similar but away from the $r$ value predicted by M1. Interaction is again non-marginal. 

In C5 we find that interaction is driven via both $\rho_{m}$ and $\rho_{d}.$ The results are provided in table V, Fig.13 - Fig.15. From the results for M2 and M3 we observe that a marginal phantom matter is the favoured DE candidate. M3 prefers a marginal non-cold DM candidate. The interaction is negligible for all the models M1-M3, but interaction coefficients can change sign for the models. This may be due to their predicted values constrained close to a zero value. This interaction has been used previously in \cite{ide3,he, c51,c52, c53,sb2}. In \cite{sb2} C5 was used in model M2. The model was constrained using 194 Supernovae 1a data, which constrained $\omega_{d}$ to -1.01, with mild positive interaction.

Table VI, and Fig.16 - Fig.18 gives the constrained parameters for function C6. Here interaction is driven by $\rho_{d}$ and $\rho'_{m}.$ From the fitted values of the interaction coefficient we see that the interaction is more significant for $\rho'_{m}$ as compared to $\rho_{d}.$ In M2 we see that fitted value of $\omega_{d}\sim -0.99$ whereas in M3 $\omega_{d}$ fits to the cosmological constant. Given that both M1 and M3 prefers CC as the preferred DE candidate we can say that C6 in general promotes interaction driven via CC. As in the previous cases we observe that M3 always prefers a marginal non-cold DM candidate.

Table VII, Fig.19 - Fig.21 gives data prescribed values for the free parameters in C7. Interaction is driven via $\rho'_{d}$ and $\rho_{m}.$ From the prescribed interaction parameter values in the table one can make out strong interaction for all models M1-M3. Further from the constrained value of the DE density parameter we observe that CC is not the preferred DE candidate. From the results of M3 we see that slightly phantom regime is non-cold DM is the preferred Dark sectors of this interacting model. 

The interaction function C8 is a combination of $\rho'_{m}$ and $\rho'_{d}.$ From table VIII, Fig.22 - Fig.24 we see that the interaction coefficients are marginal corresponding to both $\rho'_{m}$ and $\rho'_{d}$ favouring only marginal interaction for all models M1-M3. Further models M2 and M3 show a phantom drift of DE candidate. DM candidate again favours a marginal non-cold existence. Interaction C8 was first studied in \cite{c81} and later in \cite{sb3}.  In \cite{sb3} C8 was studied using model M2. They used Hubble data, BAO data and 580 Union 2.1 Supernovae data to constrain their model. Their model showed marginal negative interaction with $\omega_{d}$ constrained to around -1.095. This again is in agreement with our data. In the above literature there was only one interaction coefficient, that is the interactions $\alpha=\beta.$

From table IX, and Fig.25 - Fig.27 we see that interaction C9 closely follows the data in table I. Non-linear IDE function like C9 has not been studied closely in the literature so far. The results in table IX show that such interactions needs to be analysed in greater details. In \cite{c9} a general interaction term was used that included a product $\rho_{m},\rho_{d}$ and $\rho_{m}+\rho_{d}$ raised to arbitrary powers such that the sum of the exponents is one. C9 can be considered a special case of that interaction. However this exact form has not been previously studied in the literature.

Interaction models C3, C4, C6 or C7 has not been previously studied as considered here. In \cite{sp} a general interaction term was considered which was a linear combination of $\rho_{m},\rho_{d}$ and $\rho'_{m}.$ Both models C3 and C6 can be considered to be subclasses of the interaction considered there. In fact they study C6 in M1 as a subcategory. Their results with Cosmic Chronometer Hubble data and binned JLA data constrains $\eta$ to a mean negative value, while the coefficient $\alpha$ is constrained to a mean positive value. Comparing the results we see that our model provides less stringent constraints as compared to \cite{sp}.

In \cite{wyan} an interaction involving only $\rho'_{d}$ was studied. Interactions C4 or C7 are more general cases of the above. Significantly in both C4 or C7 we obtain negative mean interaction as has been predicted in \cite{wyan} where the interaction was proportional to $\rho'_{d}$ only, and in model M2. The data sets consisted of CMB, Cosmic Chronometers, BAO, JLA, Weak Lensing, Red shift space distortion and Local Hubble Data.

From the data in tables I-IX and figures 1-27, we  observe that the coincidence parameter $r$ is mostly constrained around 2.1-2.2 in all the models. Although $r$ was defined as the ratio of DE to DM, yet observational data always constrained it to a constant value irrespective of the dark sector matter densities. Further from C3 and C4 we observe that interaction driven by only $\rho_{m}$ and its derivative has a tendency to converge towards CC as the preferred DE candidate, while interaction driven via $\rho_{d}$ and its derivative favours a quintessence DE candidate. A constant feature in all the interaction functions is that model M3 always converged to a marginal non-cold DM candidate against the popular assumption of CDM in most of the literature on interacting models. Usually DM is divided into three categories. The Hot DM (HDM), Warm DM (WDM) and the Cold DM (CDM). In standard $\Lambda$CDM models, the usual DM candidate is the CDM, which is pressure-less, that is $p_{cdm}=0$. In fact CDM is associated with the large scale structure of the universe at scales beyond 1 Mpc. However CDM cannot explain efficiently the structures at small scales. Recent literature show that small scale observations can be explained better by WDM. Free streaming velocity for WDM candidates are intermediate of HDM and CDM, this helps to tackle some of the problems associated with CDM structure formation. HDM are relativistic candidates that were relevant only in the beginning of hot big bang, that is, during the hot radiation dominated epoch. Our numerical results in model M3 prefers the existence of a non-cold DM candidate of the WDM type. Recently the term interacting warm dark matter ``IWDM" has been used in the context of interacting non-cold dark matter in \cite{iw}. The authors of \cite{iw} used the interaction C2 between CC and non-cold DM. They analysed their model both at the background and perturbation level. Further they constrained their model using latest CMB, Lensing, BAO and Pantheon data. All of their results pointed to the existence non-cold (warm)DM candidate. According to them using non-cold DM candidate in interaction models ``could herald the beginning of a new era in cosmology." We point that models like M3 has not been hitherto used in the literature. Clearly our results in M3 are very much in agreement with the results of \cite{iw} where we obtained slightly non-cold DM interacting with marginally phantom DE or CC.

\begin{table}
\begin{tabularx}{.8\textwidth}{CCCCCCCCC}
\hhline{=========}
&Models&$\chi^{2}_{min}$&$r$&$\omega_{d}$&$\omega_{m}$&$\Omega_{d}$& $\xi$&\\
    \hline
&{\bf M1}&0.789&$2.27^{+0.50}_{-0.51}$&--&--&$0.71^{+0.02}_{-0.02}$&$0.04^{+0.03}_{-0.03}$&\\
    
&{\bf M2}&0.795&$2.24^{+0.52}_{-0.51}$&$-1.00^{+0.20}_{-0.14}$&--&$0.70^{+0.12}_{-0.08}$&$0.04^{+0.06}_{-0.12}$&\\
    
&{\bf M3}&0.81&$2.23^{+0.52}_{-0.49}$&$-1.01^{+0.21}_{-0.14}$&$0.04^{+0.04}_{-0.04}$&$0.71^{+0.12}_{-0.07}$&$0.08^{+0.07}_{-0.13}$&\\
\hhline{=========}
\end{tabularx}
\caption[One]{The modelled parameter constraints for IDE function C1 }
\end{table}

\begin{table}
\begin{tabularx}{.8\textwidth}{CCCCCCCCC}
\hhline{=========}
&Models&$\chi^{2}_{min}$&$r$&$\omega_{d}$&$\omega_{m}$&$\Omega_{d}$& $\eta$&\\
    \hline
&{\bf M1}&0.789&$2.17^{+0.53}_{-0.47}$&--&--&$0.71^{+0.02}_{-0.02}$&$0.02^{+0.01}_{-0.01}$&\\
    
&{\bf M2}&0.795&$2.14^{+0.55}_{-0.46}$&$-1.01^{+0.19}_{-0.13}$&--&$0.70^{+0.11}_{-0.07}$&$0.02^{+0.03}_{-0.05}$&\\
    
&{\bf M3}&0.81&$2.14^{+0.56}_{-0.47}$&$-1.02^{+0.19}_{-0.13}$&$0.04^{+0.04}_{-0.04}$&$0.71^{+0.11}_{-0.07}$&$0.04^{+0.03}_{-0.05}$&\\
\hhline{=========}
\end{tabularx}
\caption[Two]{The modelled parameter constraints for IDE function C2 }
\end{table}

\begin{table}
\begin{tabularx}{.9\textwidth}{CCCCCCCCCC}
\hhline{==========}
&Models&$\chi^{2}_{min}$&$r$&$\omega_{d}$&$\omega_{m}$&$\Omega_{d}$& $\xi$&$\alpha$&\\
    \hline
&{\bf M1}&0.795&$2.27^{+0.51}_{-0.51}$&--&--&$0.67^{+0.08}_{-0.08}$&$-0.08^{+0.24}_{-0.17}$&$-0.16^{+0.31}_{-0.24}$&\\
    
&{\bf M2}&0.807&$2.26^{+0.51}_{-0.51}$&$-1.00^{+0.19}_{-0.14}$&--&$0.67^{+0.11}_{-0.09}$&$-0.10^{+0.27}_{-0.19}$&$-0.18^{+0.34}_{-0.22}$ &\\
  
&{\bf M3}&0.82&$2.26^{+0.50}_{-0.51}$&$-1.00^{+0.19}_{-0.13}$&$0.04^{+0.04}_{-0.04}$&$0.69^{+0.11}_{-0.10}$&$-0.07^{+0.28}_{-0.20}$&$-0.17^{+0.33}_{-0.28}$&\\

\hhline{==========}
\end{tabularx}
\caption[Three]{The modelled parameter constraints for IDE function C3 }
\end{table}

\begin{table}
\begin{tabularx}{.9\textwidth}{CCCCCCCCCC}
\hhline{==========}
&Models&$\chi^{2}_{min}$&$r$&$\omega_{d}$&$\omega_{m}$&$\Omega_{d}$& $\eta$&$\beta$&\\
    \hline
&{\bf M1}&0.795&$2.20^{+0.54}_{-0.51}$&--&--&$0.59^{+0.13}_{-0.10}$&$-0.13^{+0.18}_{-0.11}$&$-0.22^{+0.25}_{-0.19}$&\\
    
&{\bf M2}&0.807&$2.11^{+0.58}_{-0.44}$&$-0.98^{+0.18}_{-0.15}$&--&$0.65^{+0.13}_{-0.11}$&$-0.17^{+0.23}_{-0.16}$&$-0.23^{+0.28}_{-0.18}$ &\\ 
  
&{\bf M3}&0.82&$2.15^{+0.52}_{-0.45}$&$-0.92^{+0.23}_{-0.19}$&$0.05^{+0.04}_{-0.04}$&$0.64^{+0.14}_{-0.12}$&$-0.17^{+0.20}_{-0.13}$&$-0.26^{+0.25}_{-0.17}$&\\

\hhline{==========}
\end{tabularx}
\caption[Four]{The modelled parameter constraints for IDE function C4 }
\end{table}

\begin{table}
\begin{tabularx}{.9\textwidth}{CCCCCCCCCC}
\hhline{==========}
&Models&$\chi^{2}_{min}$&$r$&$\omega_{d}$&$\omega_{m}$&$\Omega_{d}$& $\xi$&$\eta$&\\
    \hline
&{\bf M1}&0.802&$2.09^{+0.59}_{-0.42}$&--&--&$0.71^{+0.02}_{-0.02}$&$-0.05^{+0.37}_{-0.32}$&$0.04^{+0.15}_{-0.17}$&\\
    
&{\bf M2}&0.807&$2.15^{+0.53}_{-0.46}$&$-1.01^{+0.20}_{-0.13}$&--&$0.70^{+0.12}_{-0.07}$&$0.03^{+0.31}_{-0.36}$&$-0.00^{+0.16}_{-0.15}$&\\ 
 
&{\bf M3}&0.82&$2.14^{+0.57}_{-0.45}$&$-1.02^{+0.23}_{-0.19}$&$0.04^{+0.04}_{-0.04}$&$0.70^{+0.12}_{-0.07}$&$-0.01^{+0.33}_{-0.35}$&$-0.02^{+0.17}_{-0.16}$&\\

\hhline{==========}
\end{tabularx}
\caption[Five]{The modelled parameter constraints for IDE function C5 }
\end{table}

\begin{table}
\begin{tabularx}{.9\textwidth}{CCCCCCCCCC}
\hhline{==========}
&Models&$\chi^{2}_{min}$&$r$&$\omega_{d}$&$\omega_{m}$&$\Omega_{d}$& $\eta$&$\alpha$&\\
    \hline
&{\bf M1}&0.795&$2.17^{+0.52}_{-0.48}$&--&--&$0.67^{+0.08}_{-0.07}$&$-0.03^{+0.12}_{-0.08}$&$-0.15^{+0.34}_{-0.25}$&\\
   
&{\bf M2}&0.807&$2.10^{+0.59}_{-0.43}$&$-0.99^{+0.21}_{-0.15}$&--&$0.69^{+0.12}_{-0.10}$&$-0.05^{+0.14}_{-0.10}$&$-0.17^{+0.37}_{-0.24}$&\\ 

&{\bf M3}&0.82&$2.17^{+0.55}_{-0.48}$&$-1.00^{+0.21}_{-0.14}$&$0.04^{+0.04}_{-0.04}$&$0.69^{+0.11}_{-0.10}$&$-0.03^{+0.13}_{-0.10}$&$-0.17^{+0.35}_{-0.24}$&\\

\hhline{==========}
\end{tabularx}
\caption[Six]{The modelled parameter constraints for IDE function C6 }
\end{table}

\begin{table}
\begin{tabularx}{.9\textwidth}{CCCCCCCCCC}
\hhline{==========}
&Models&$\chi^{2}_{min}$&$r$&$\omega_{d}$&$\omega_{m}$&$\Omega_{d}$& $\xi$&$\beta$&\\
    \hline
&{\bf M1}&0.795&$2.18^{+0.52}_{-0.47}$&--&--&$0.62^{+0.10}_{-0.08}$&$-0.21^{+0.30}_{-0.20}$&$-0.16^{+0.19}_{-0.15}$&\\
   
&{\bf M2}&0.807&$2.14^{+0.55}_{-0.42}$&$-1.00^{+0.17}_{-0.14}$&--&$0.63^{+0.13}_{-0.10}$&$-0.20^{+0.32}_{-0.21}$&$-0.14^{+0.19}_{-0.15}$&\\ 

&{\bf M3}&0.82&$2.14^{+0.55}_{-0.46}$&$-1.01^{+0.18}_{-0.13}$&$0.05^{+0.04}_{-0.04}$&$0.64^{+0.18}_{-0.12}$&$-0.18^{+0.33}_{-0.22}$&$-0.15^{+0.19}_{-0.17}$&\\

\hhline{==========}
\end{tabularx}
\caption[Seven]{The modelled parameter constraints for IDE function C7 }
\end{table}

\begin{table}
\begin{tabularx}{.9\textwidth}{CCCCCCCCCC}
\hhline{==========}
&Models&$\chi^{2}_{min}$&$r$&$\omega_{d}$&$\omega_{m}$&$\Omega_{d}$& $\alpha$&$\beta$&\\
    \hline
&{\bf M1}&0.801&$2.07^{+0.60}_{-0.41}$&--&--&$0.69^{+0.02}_{-0.03}$&$-0.05^{+0.36}_{-0.31}$&$-0.00^{+0.15}_{-0.17}$&\\

&{\bf M2}&0.807&$2.18^{+0.54}_{-0.46}$&$-1.04^{+0.16}_{-0.11}$&--&$0.66^{+0.12}_{-0.07}$&$-0.02^{+0.33}_{-0.35}$&$-0.04^{+0.16}_{-0.15}$&\\ 

&{\bf M3}&0.82&$2.11^{+0.59}_{-0.43}$&$-1.03^{+0.18}_{-0.13}$&$0.05^{+0.03}_{-0.04}$&$0.66^{+0.14}_{-0.08}$&$-0.00^{+0.32}_{-0.34}$&$-0.04^{+0.15}_{-0.17}$&\\

\hhline{==========}
\end{tabularx}
\caption[Eight]{The modelled parameter constraints for IDE function C8 }
\end{table}

\begin{table}
\begin{tabularx}{.9\textwidth}{CCCCCCCCCC}
\hhline{==========}
&Models&$\chi^{2}_{min}$&$r$&$\omega_{d}$&$\omega_{m}$&$\Omega_{d}$& $\gamma$&$\delta$&\\
    \hline
&{\bf M1}&0.801&$2.14^{+0.54}_{-0.46}$&--&--&$0.70^{+0.02}_{-0.02}$&$0.02^{+0.02}_{-0.02}$&$0.09^{+0.30}_{-0.37}$&\\

&{\bf M2}&0.808&$2.17^{+0.57}_{-0.47}$&$-1.00^{+0.20}_{-0.14}$&--&$0.70^{+0.12}_{-0.07}$&$0.02^{+0.03}_{-0.06}$&$0.08^{+0.31}_{-0.36}$&\\ 

&{\bf M3}&0.82&$2.17^{+0.55}_{-0.46}$&$-1.03^{+0.18}_{-0.12}$&$0.04^{+0.04}_{-0.04}$&$0.69^{+0.11}_{-0.07}$&$0.04^{+0.04}_{-0.05}$&$0.09^{+0.30}_{-0.38}$&\\

\hhline{==========}
\end{tabularx}
\caption[Nine]{The modelled parameter constraints for IDE function C9 }
\end{table}

\section{Effective Equation of State}

Here we evaluate the effective equation of state $\omega_{t}=\frac{\rho_{m}\omega_{m}+\rho_{d}\omega_{d}}{\rho_{m}+\rho_{d}}.$ Using equations (\ref{rm}) and (\ref{rd}) we write $\omega_{t}=\frac{\omega_{m}\rho_{m0}(1+z)^{3m_{1}}+\omega_{d}\rho_{d0}(1+z)^{3m_{2}}}{\rho_{m0}(1+z)^{3m_{1}}+\rho_{d0}(1+z)^{3m_{2}}},$ where $m_{1}=\frac{1+\omega_{m}-\xi-\eta r-\gamma r^{1-\delta}}{1-\alpha-\beta r}$ and $m_{2}=\frac{1+\omega_{d}+\frac{\xi}{r}+\eta+\frac{\gamma}{r^{\delta}}}{1+\frac{\alpha}{r}+\beta}.$ From this we see that, 
\begin{itemize}
\item as $z\rightarrow\infty,~\omega_{t}\rightarrow\begin{cases}
\omega_{d}, \text{if}~m_{2}-m_{1}>0,\\
\omega_{m}, \text{if}~m_{2}-m_{1}<0
\end{cases},$ 
\item as $z\rightarrow -1,~\omega_{t}\rightarrow\begin{cases}
\omega_{m}, \text{if}~m_{2}-m_{1}>0,\\
\omega_{d}, \text{if}~m_{2}-m_{1}<0
\end{cases}.$ 
\end{itemize}
We see in that for all the interaction functions C1-C9 and Models M1-M3, $\omega_{t}\rightarrow\omega_{m}$ as $z\rightarrow\infty$ while $\omega_{t}\rightarrow\omega_{d}$ as $z\rightarrow -1.$ This shows that all the models predict an early matter dominated epoch followed by a late time DE dominated epoch. It may be noted that, it is only in the recent epoch that DE and DM have evolved to similar order of magnitude. However such constant $r$ is not valid in an early universe. Yet, once we have deduced the model using analytical as well as numerical means, it is imperative that we check the model behaviour in the limiting scenarios of early and late epochs, where an early epoch would be given by a large $z$ value or in the limit $z\rightarrow\infty$ and a late or future epoch would mean $z\rightarrow-1.$ Accordingly the above limits on $z$ becomes important. Significantly the numerical results obtained above, when combined with the limiting analytical scenarios, our models provide evidence of expected early time as well as late time cosmology. 

\section{The Deceleration Factor}

The deceleration factor is defined as $q=-1-\frac{\dot{H}}{H^{2}}.$ For the given model using equation(\ref{expan1}) we can rewrite the deceleration factor as $q=-1+\frac{3}{2}\frac{m_{1}\rho_{m0}(1+z)^{3m_{1}}+m_{2}\rho_{d0}(1+z)^{3m_{2}}}{\rho_{m0}(1+z)^{3m_{1}}+\rho_{d0}(1+z)^{3m_{2}}}.$ Here also we can extrapolate the model to the limiting scenarios of $z\rightarrow\infty$ and $z\rightarrow-1$ observe the model predictions at early and late times.
\begin{itemize}
\item as $z\rightarrow\infty,~q\rightarrow\begin{cases}
-1+\frac{3}{2}m_{2}, \text{if}~m_{2}-m_{1}>0,\\
-1+\frac{3}{2}m_{1}, \text{if}~m_{2}-m_{1}<0
\end{cases},$ 
\item as $z\rightarrow -1,~q\rightarrow\begin{cases}
-1+\frac{3}{2}m_{1}, \text{if}~m_{2}-m_{1}>0,\\
-1+\frac{3}{2}m_{2}, \text{if}~m_{2}-m_{1}<0
\end{cases}.$ 
\end{itemize}
Analysing this for all our models we find that $q>0$ as $z\rightarrow\infty$ while $q<0$ as $z\rightarrow -1.$ This is expected with models predicting an early time matter dominated decelerating universe followed by a late time accelerating universe.

\section{Model Selection based on information criteria}

With the introduction of data based research in cosmology, models have taken a centre stage in the efforts to understand the universe. Accordingly prioritising one model over the other has become an important aspect in cosmology. The most commonly used model selection criterion are the Akaike Information Criteria (AIC) \cite{ak} and the Bayesian Information Criteria (BIC) (sometimes called the Schwarz information criteria) \cite{bk}. Both these criteria penalizes a model based on the complexity introduced by the extra set of parameters. A model with more parameters is likely to have an improved likelihood estimation as compared with one with less number of parameters. However this does not always give us a complete picture. AIC and BIC both compensates for improved likelihood by correcting on the number of parameters. The AIC directly compensates on the number of parameters by $-2\ln \mathfrak{L}+2d$ where $d$ is the total number of free parameters, while BIC is defined as $-2\ln \mathfrak{L}+d\ln N$ where $N$ is the total number of data points used. The usual way to use these two information criteria is not by actual estimation but by comparing it with a base model with minimum number of parameters. AIC is said to be the criteria that defines how well a model is favoured as compared to the base model, while BIC defines how well a model is rejected as compared to the base model.  $0<\Delta AIC<2$ can be interpreted as a strongly favoured model, while $4<\Delta AIC<7$ means some evidence in favour of the model. $\Delta AIC >10$ stands for no evidence in favour. $0<\Delta BIC<2$ can be interpreted as not much strong evidence to reject the model, $2<\Delta BIC<6$ can be interpreted as some evidence against the model, while $\Delta BIC>10$ means good evidence to reject the model. 

Since all our interacting coefficients were treated under three different models, with M1 containing the least number of parameters, we consider the AIC and BIC values for M1 in the corresponding interactions C1-C9 as the base model and compare it with M2 and M3. Our results as listed in table X, show that for all the interactions C1-C9, model M2 is highly favoured by AIC although by BIC we can get some evidence against it. AIC is less harsh with M3 as is expected, which says that the model somewhat favoured in comparison with M1. Whereas BIC being more stringent is more harsh with M3. However none of the models can be ruled out in comparison to M1. 

If $\Lambda$CDM is chosen as a base model for M1 and $\omega$CDM as a base model for M2 and M3, then both M1 and M2 perform well, with IC values for M1 in the models C1-C9 ranging between $1.4\leq\Delta AIC_{\Lambda}<3,~3<\Delta BIC_{\Lambda}<8,$ and for M2 the IC values for all the functions range between  $2<\Delta AIC_{\omega}<3,~3<\Delta BIC_{\omega}<8.$ However in case of M3, when models C1-C9 is compared with $\omega$CDM we obtain the following range for IC vales: $3<\Delta AIC_{\omega}<5.5,~7.5<\Delta BIC_{\omega}<12.1$ This shows that BIC hints rejection of the model M3 in comparison to $\omega$CDM as base model. 
\begin{table}
\centering
\begin{tabularx}{\textwidth}{C|CCCCC}
\hline
Interaction&Model&AIC&BIC&$\Delta$AIC&$\Delta$BIC\\
\hline
\multirow{3}{*}{C1-C2}&M1&$\simeq 58.91$&$\simeq65.65$&-&-\\
\cline{2-6}
&M2&$\simeq 60.51$&$\simeq 69.50$&1.6&3.85\\
\cline{2-6}
&M3&$\simeq 62.5$&$\simeq 73.75$&3.6&8.1\\
\hhline{======}
\multirow{3}{*}{C3-C8}&M1&$\simeq 60.5$&$\simeq 69.49$&-&-\\
\cline{2-6}
&M2&$\simeq 62.49$&$\simeq 73.73$&1.98&4.24\\
\cline{2-6}
&M3&$\simeq 64.49$&$\simeq 77.98$&3.99&8.49\\
\hhline{======}
\multirow{3}{*}{C9}&M1&60.87&69.86&-&-\\
\cline{2-6}
&M2&62.52&73.76&1.65&3.9\\
\cline{2-6}
&M3&64.51&78&3.64&8.14\\
\hhline{======}

\end{tabularx}
\caption[Ten]{The AIC and BIC value with corresponding $\Delta$IC value of models M2 and M3 when compared with M1 as a base models}
\end{table}

\section{Discussion}

In this article we have considered a scenario of two interacting fluids. Usually such models have been analysed under the purview of interacting CDM and vacuum energy or the CDM and DE. Here we have broadened our perspective to generalise DM and have considered a third model of interacting ``two fluids". These three models were then analysed using nine interacting functions, all obtained from a general interaction term. 

In this article the main purpose was to bring forth a comparative study of several competing interacting models and interacting functions under one umbrella. Here we have also introduced the coincidence factor $r$ as a parameter of the model. We have further computed the effective equation of state as well as the deceleration parameter for the models considered. Statistical model selection analysis shows relatively better acceptance for models M1 and M2 as compared to M3. This is however expected given that, the AIC and BIC tests are based on comparison w.r.t. already existing simple models. Here M3 had no base model to compare unlike M1 which has $\Lambda$CDM  and M2 which had the $\omega$CDM as their base model for comparison. However based on comparison with M1 as base model M3 could not be completely rejected. The numerical results showed that all models favoured mostly vacuum DE or marginally phantom DE, with M3 additionally preferring a slightly non-cold DM candidate. Interaction function C4, which is driven via $\rho_{d}$ and $\rho'_{d}$ however preferred a quintessence DE candidate in both M3 and M2. This slightly difference in choice of C4 needs further analysis. Clearly the results of C4 from M1 shows that CC is not a favoured DE candidate for this interaction. This result along with the fact that when DM equation of state is left free to evolve has a tendency to develop ``non-cold" nature is something that needs further analysis. 

The current work offers only a glimpse and overview of the possibilities of information that the models can offer. The interaction functions and the models studied in this article needs greater subjective analysis using a comprehensive pool of data sets, which we aim to do in near future. Further one also needs to understand the stability of the above interaction functions via a perturbation analysis, which will help us understand the underlying mechanics of the factors driving the interactions. Preliminary analysis raises interesting possibilities and show that the above interacting functions are viable choices for interacting two fluid scenario of the universe. One may note note that models like M3 which have not been extensively studied in the literature can lead to interesting and otherwise unexplored possibilities. In this context, results of \cite{iw} increases curiosity in the context of M3. In \cite{iw} C2 was treated from the perspective of interacting vacuum energy with a perfect fluid equation of state, which the authors called interacting warm dark matter and obtained similar results for the choice of DM candidate. This clearly indicates that there is a need to understand and study models as M3 with greater depth and understanding.

\section{ Acknowledgemensts:}

SB acknowledges the Department of Science and Technology, SERB, India for financial help through ECR project (File No. ECR/2017/000569). 
SB also acknowledges IUCAA, Pune, India, where the project was initiated.

The author acknowledges the anonymous reviewers whose feedback helped to improve the    article.

\appendix

\section{Figures}

The following figures show the confidence contour on a two dimensional parameter space with marginalized posterior distributions corresponding to the interaction C1-C9 and models M1-M3. The figures were obtained by a combined analysis of binned-LA+OHD+BAO data.The median values of the constrained parameters with corresponding $1\sigma$ errors are given on top of the posterior probability distribution panels.

\begin{figure}[htb]
\begin{center}
\includegraphics[scale=0.5]{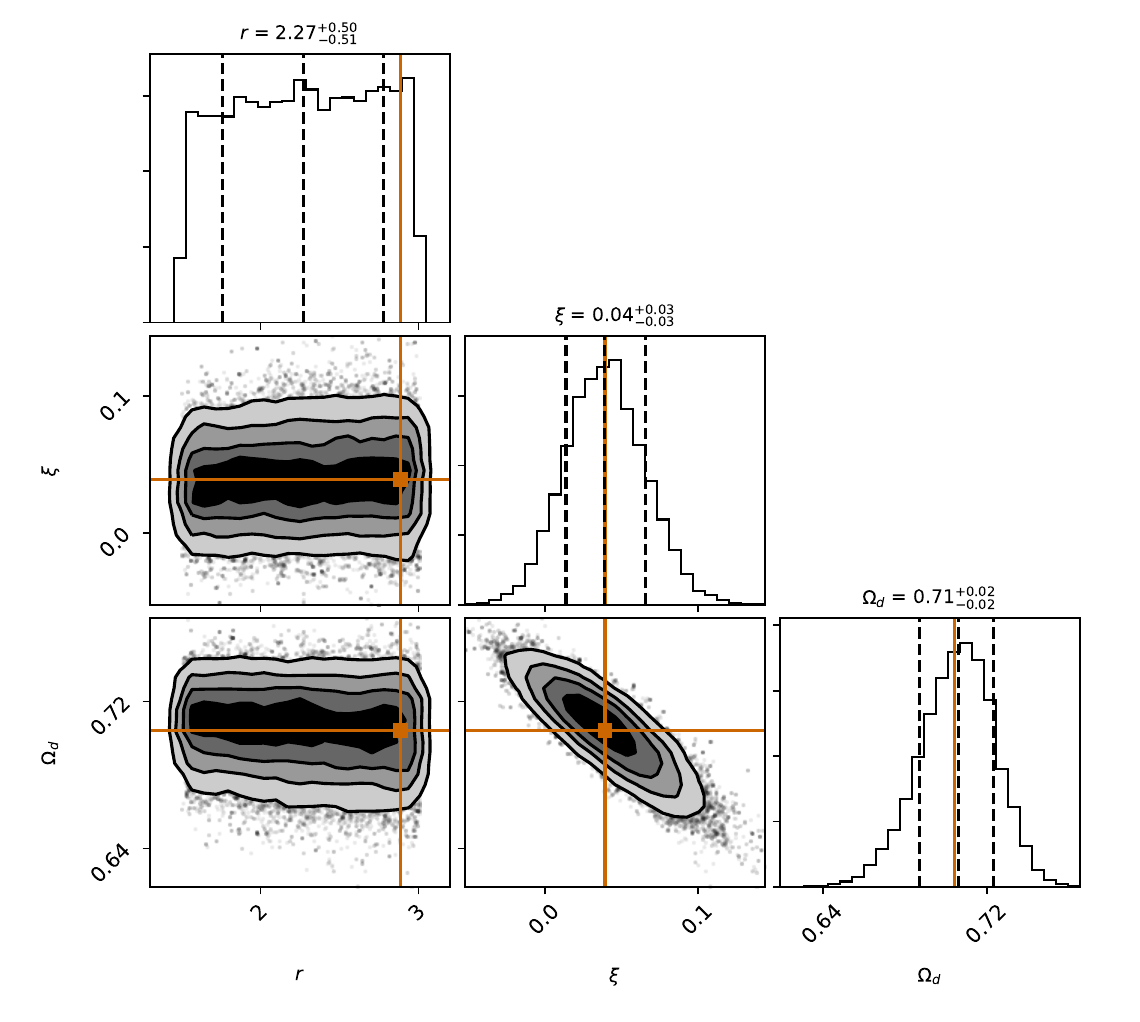}\quad
\caption{The above depicts the confidence contour for function C1 and model M1.}
\end{center}
\end{figure}

\begin{figure}[htb]
\begin{center}
\includegraphics[scale=0.5]{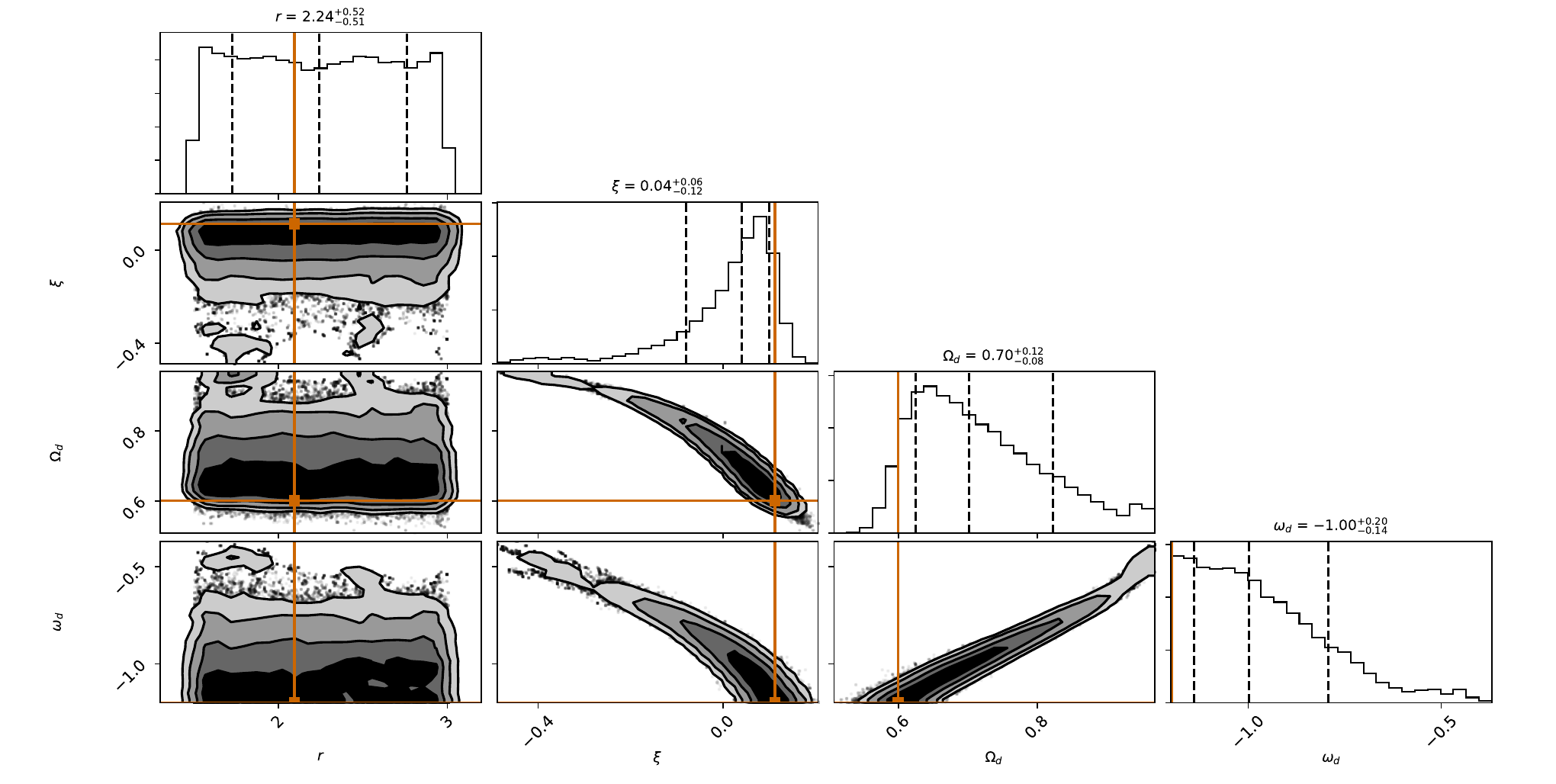}\quad
\caption{The above depicts the confidence contour for function C1 and model M2.}
\end{center}
\end{figure}

\begin{figure}[htb]
\begin{center}
\includegraphics[scale=0.5]{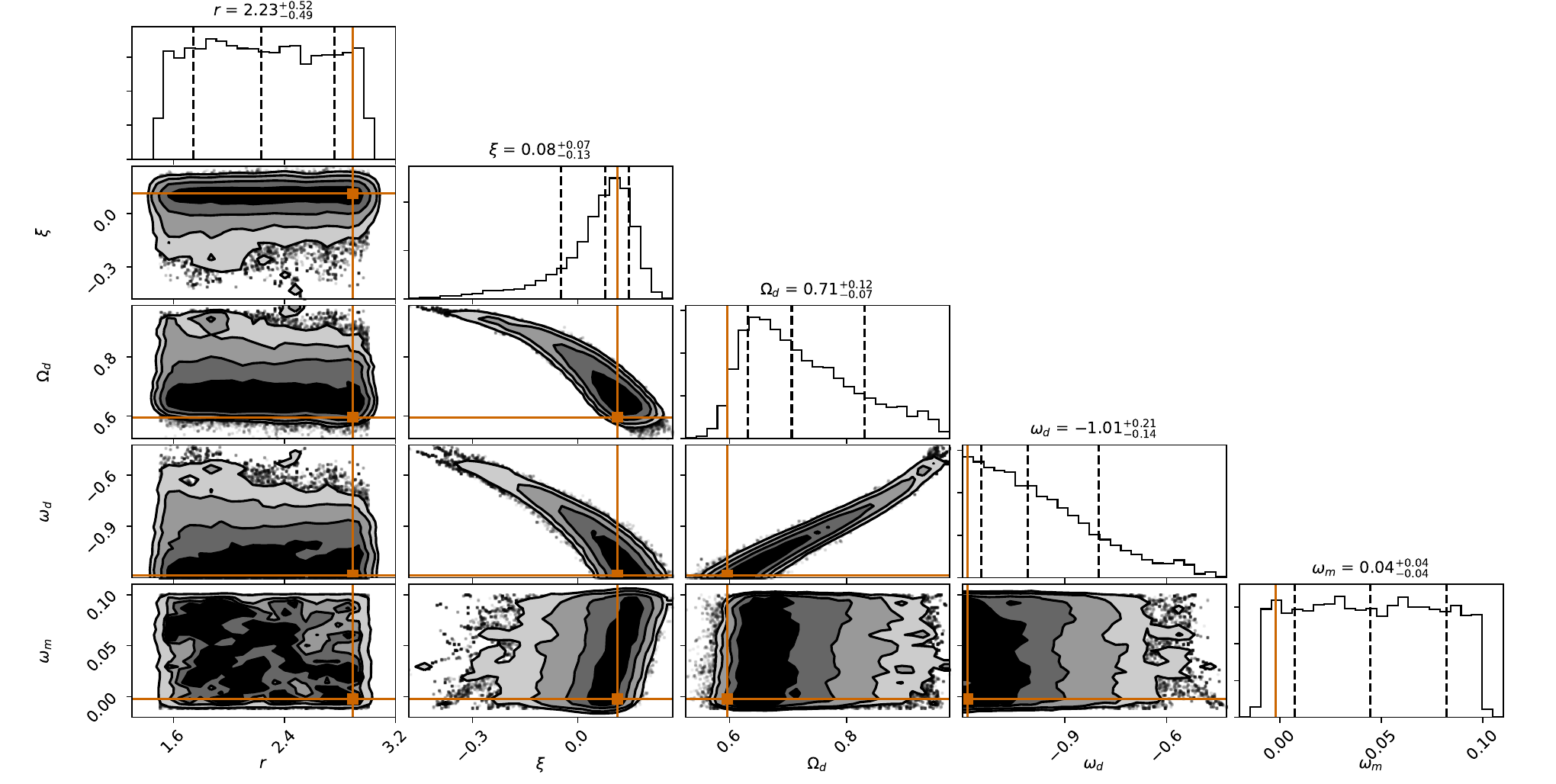}
\caption{The above depicts the confidence contour for function C1 and model M3.}
\end{center}
\end{figure}

\begin{figure}[htb]
\begin{center}
\includegraphics[scale=0.5]{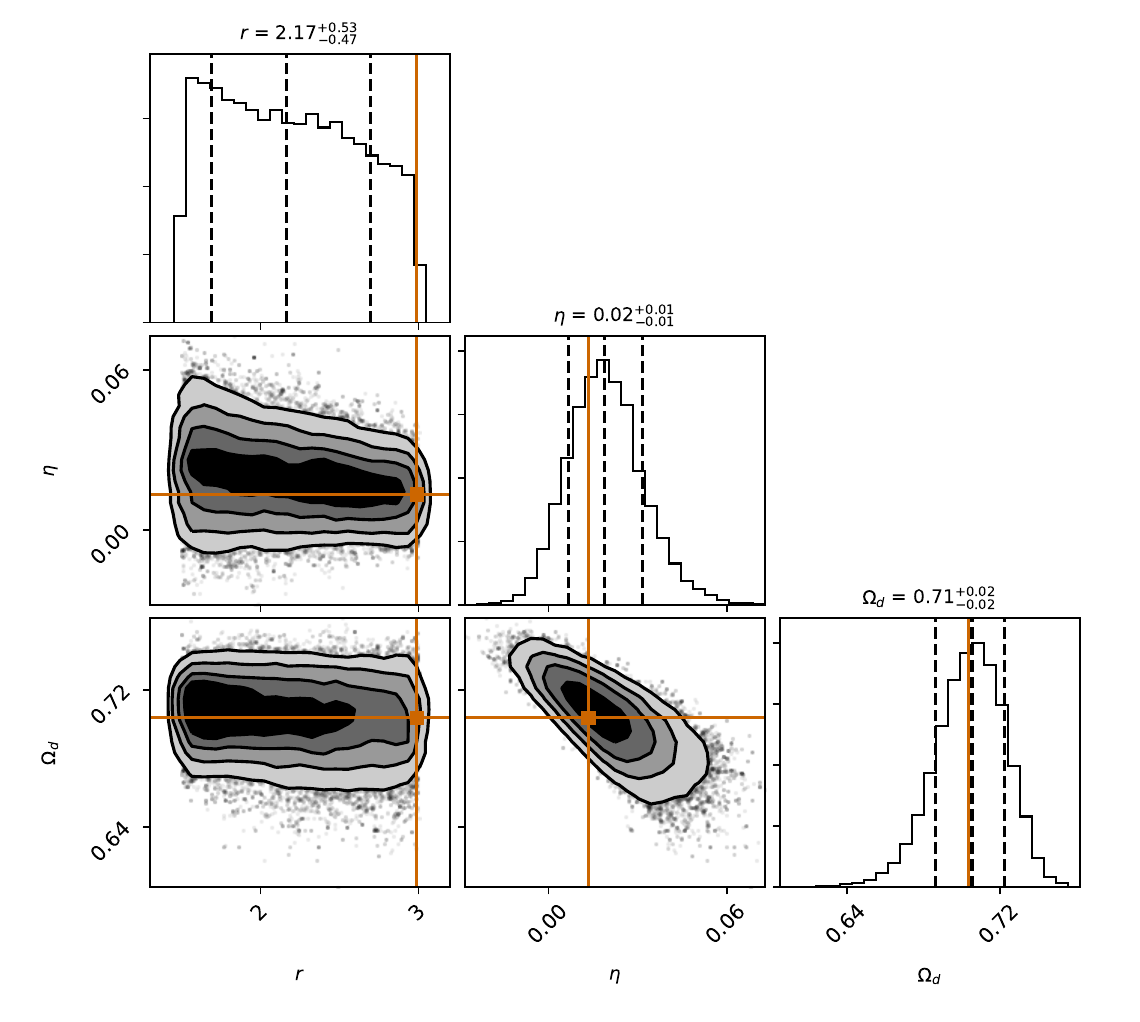}\quad
\caption{The above depicts the confidence contour for function C2 and model M1.}
\end{center}
\end{figure}

\begin{figure}[htb]
\begin{center}
\includegraphics[scale=0.5]{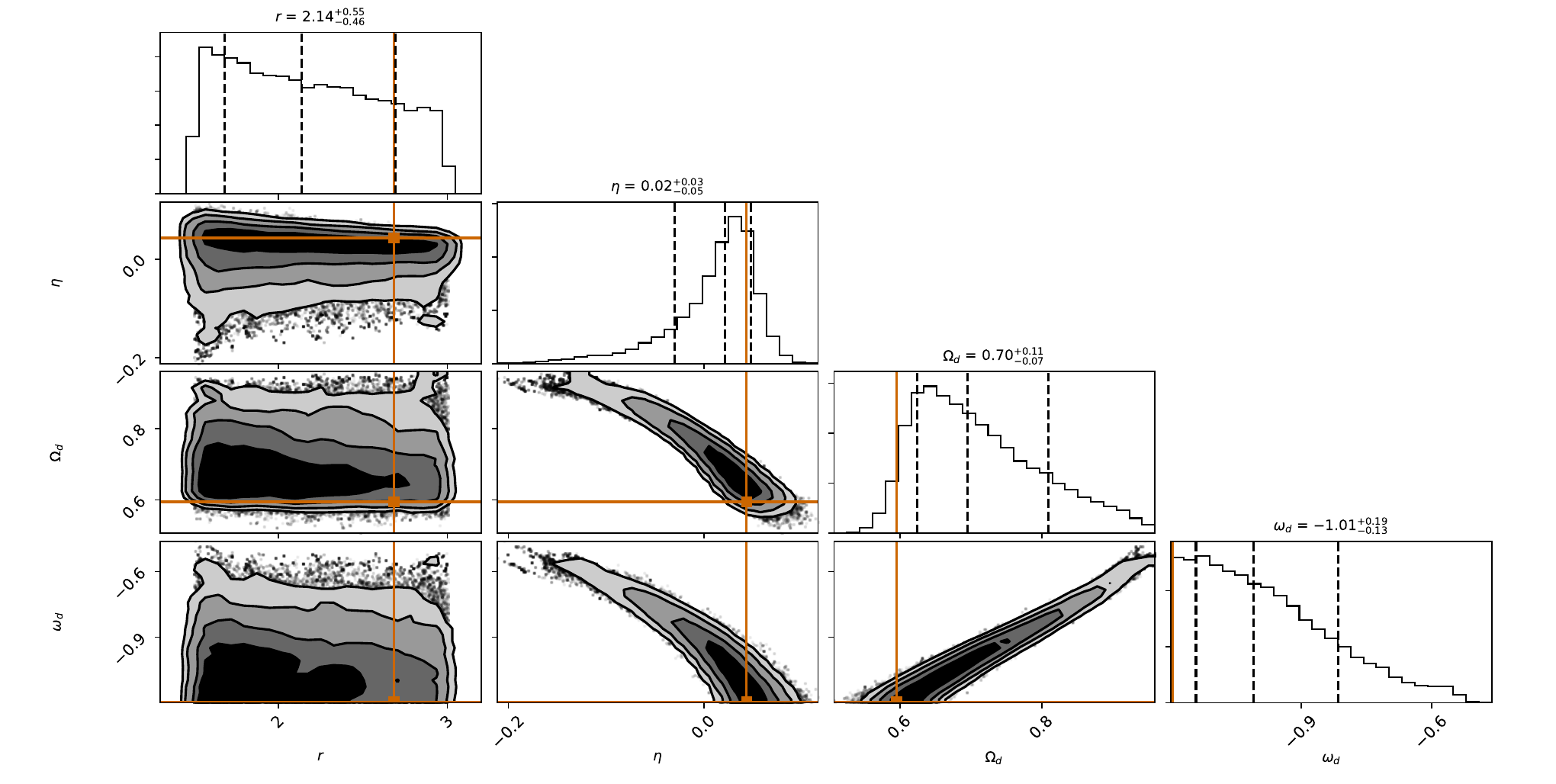}
\caption{The above depicts the confidence contour for function C2 and model M2.}
\end{center}
\end{figure}

\begin{figure}[htb]
\begin{center}
\includegraphics[scale=0.5]{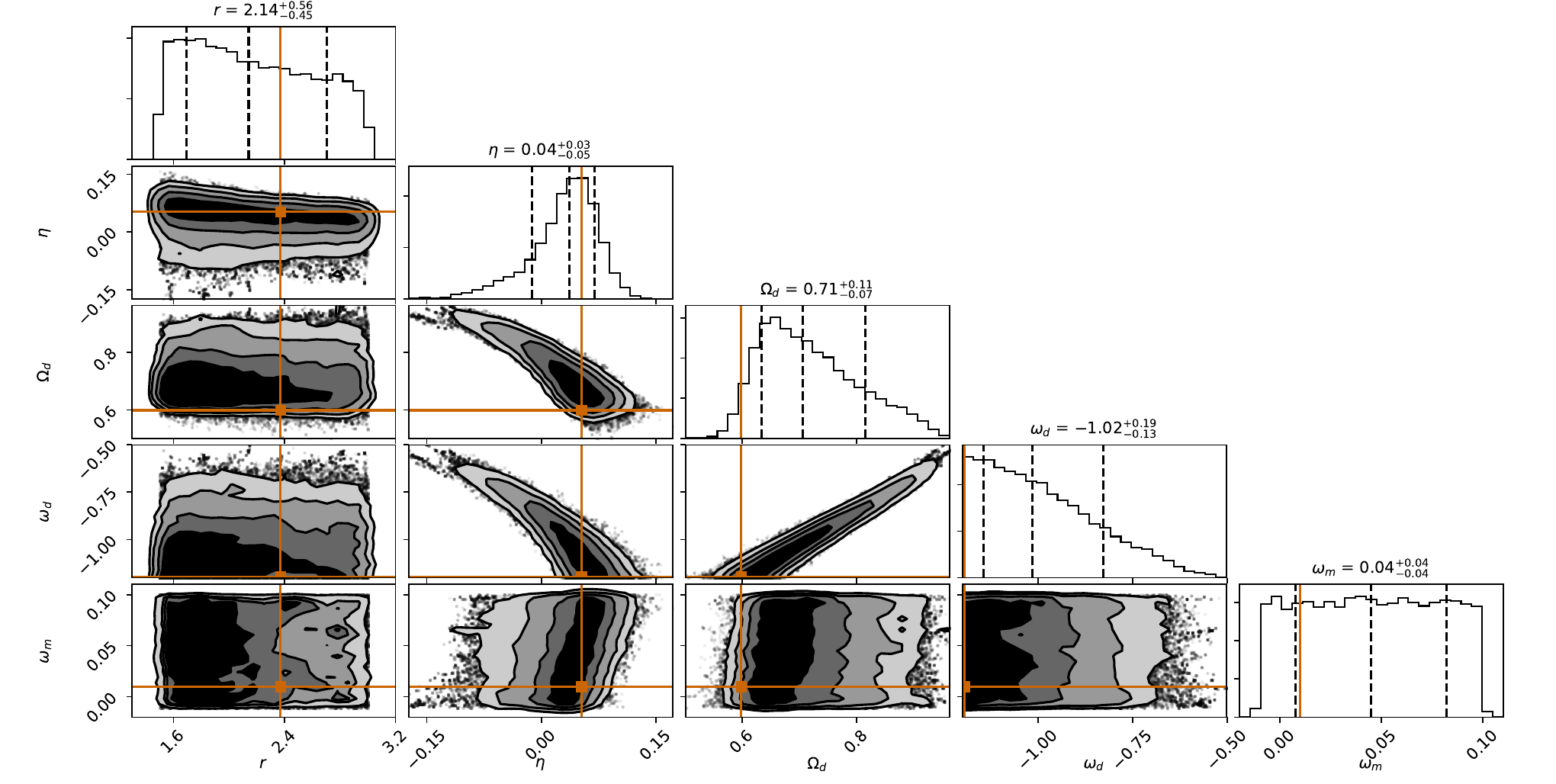}
\caption{The above depicts the confidence contour for function C2 and model M3.}
\end{center}
\end{figure}

\begin{figure}[htb]
\begin{center}
\includegraphics[scale=0.5]{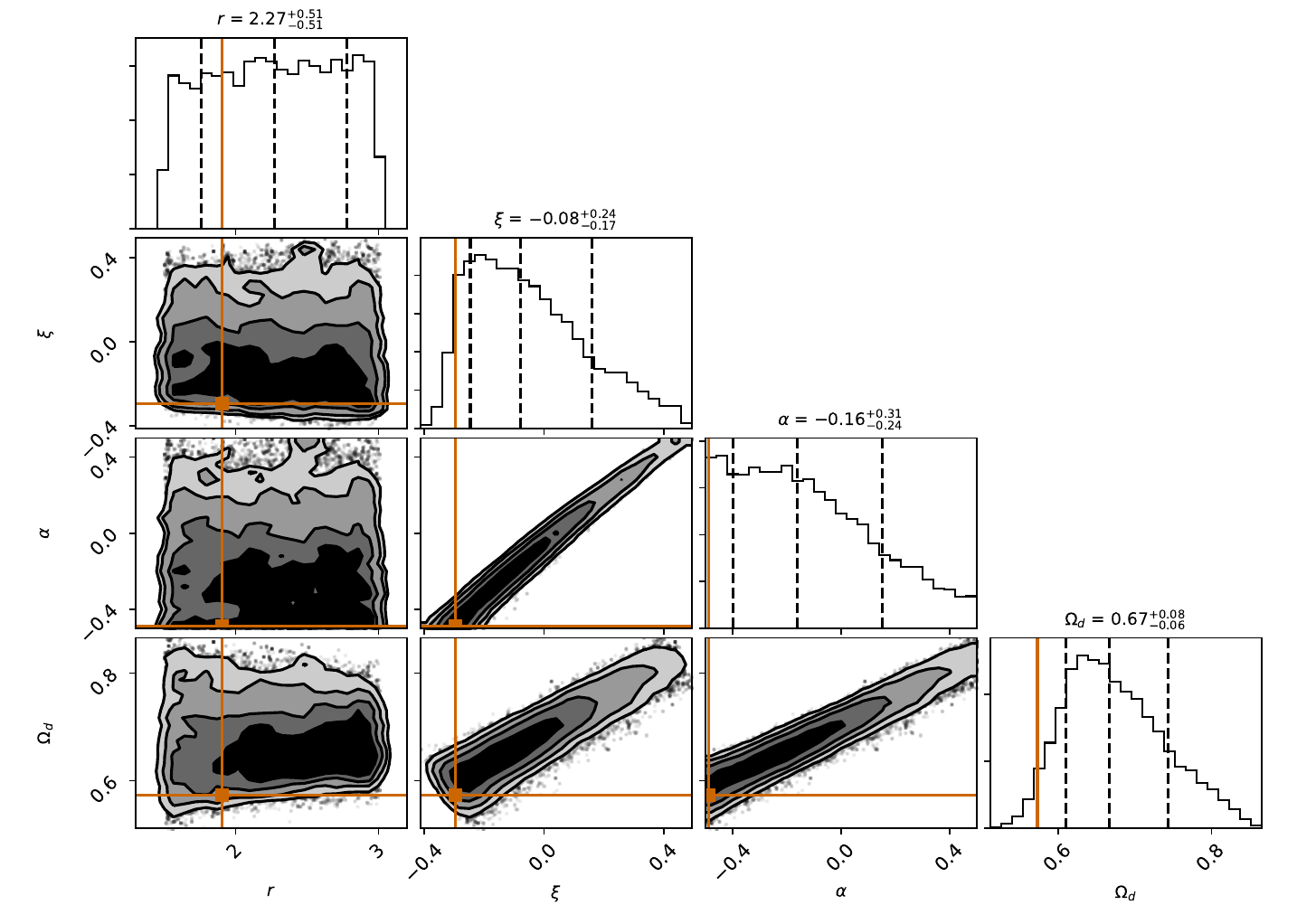}
\caption{The above depicts the confidence contour for function C3 and model M1.}
\end{center}
\end{figure}

\begin{figure}[htb]
\begin{center}
 \includegraphics[scale=0.5]{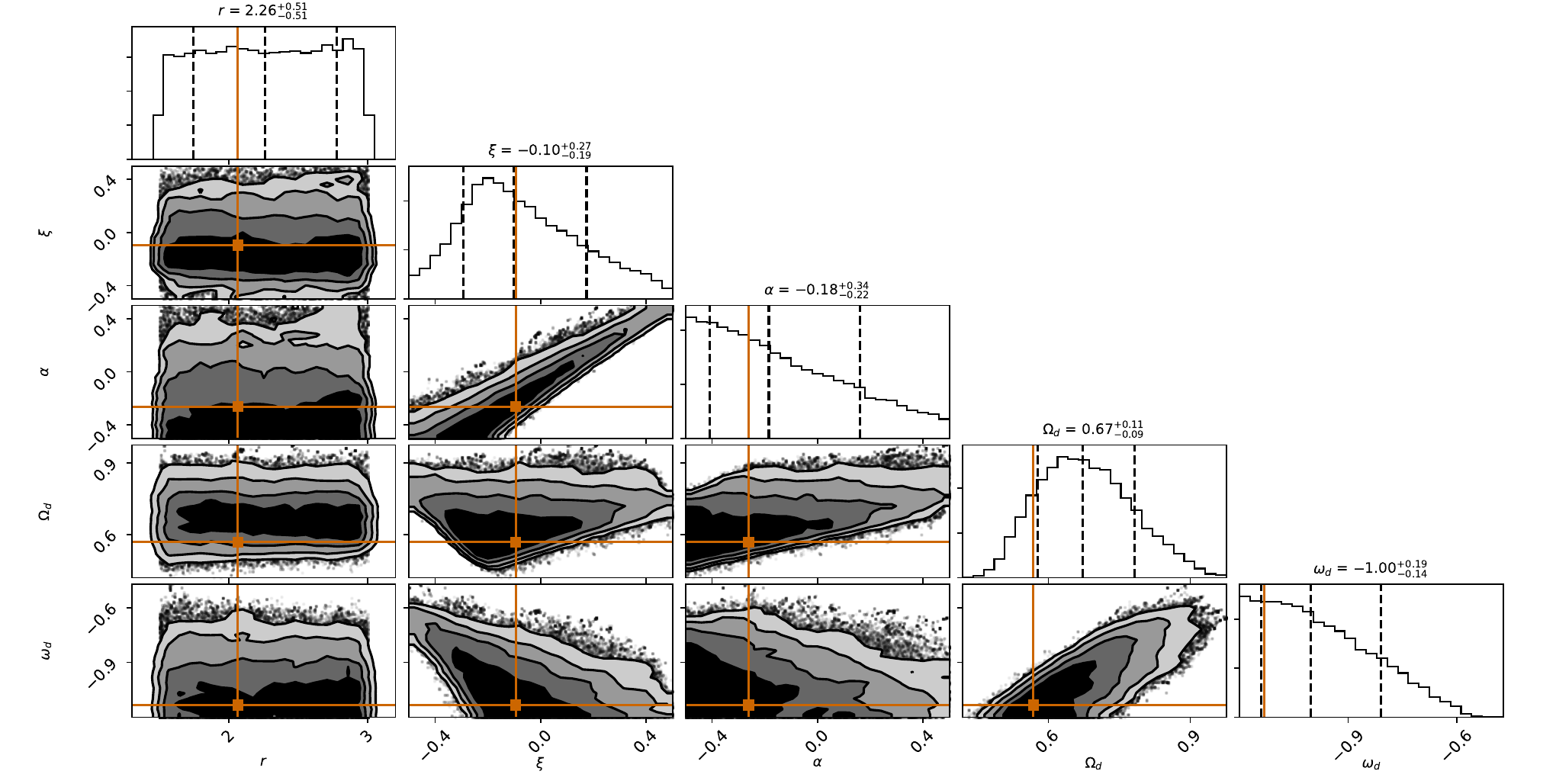}
\caption{The above depicts the confidence contour for function C3 and model M2.}
\end{center}
\end{figure}

\begin{figure}[htb]
\begin{center}
\includegraphics[scale=0.5]{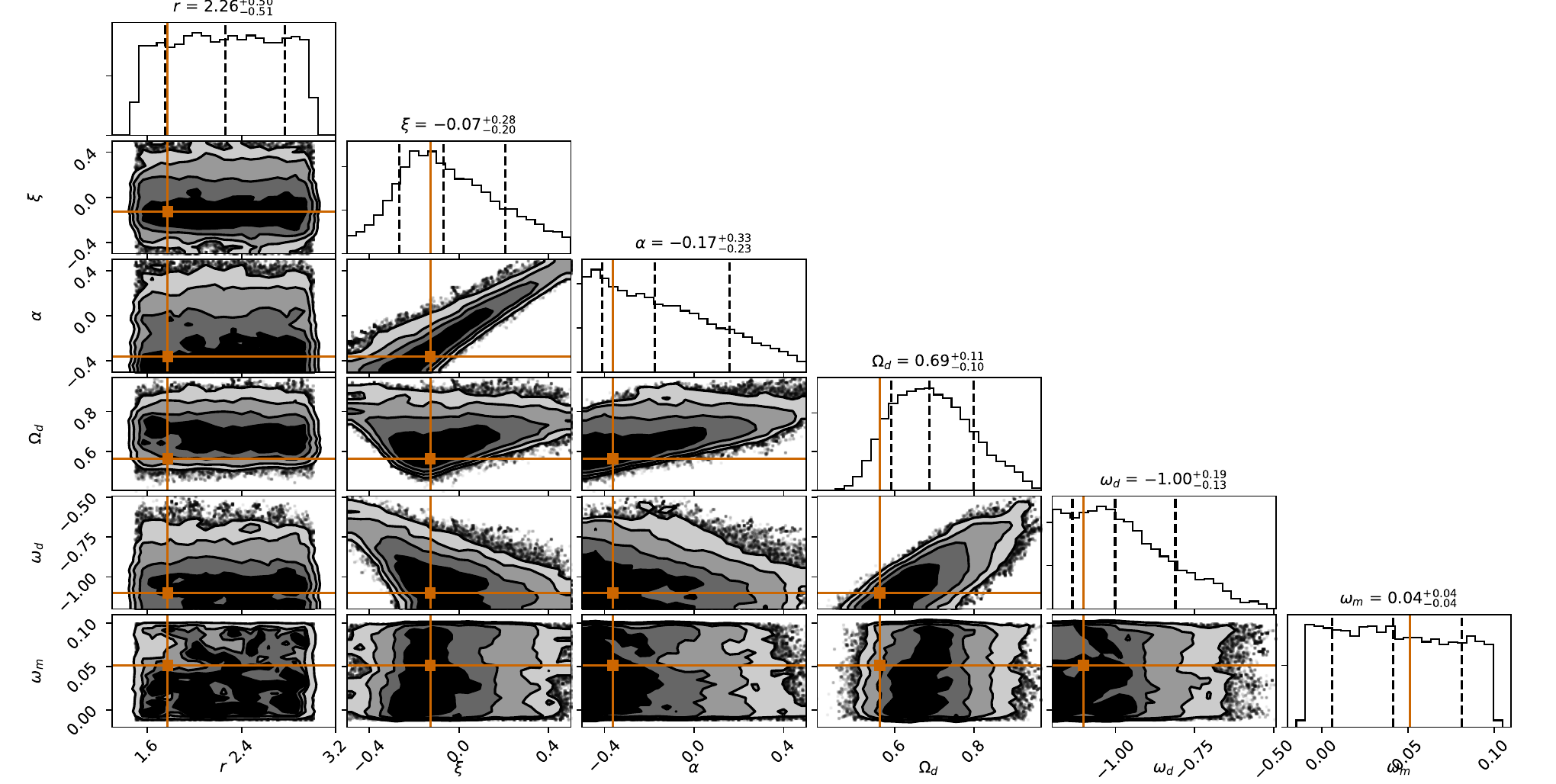}
\caption{The above depicts the confidence contour for function C3 and model M3.}
\end{center}
\end{figure}

\begin{figure}[htb]
\begin{center}
\includegraphics[scale=0.5]{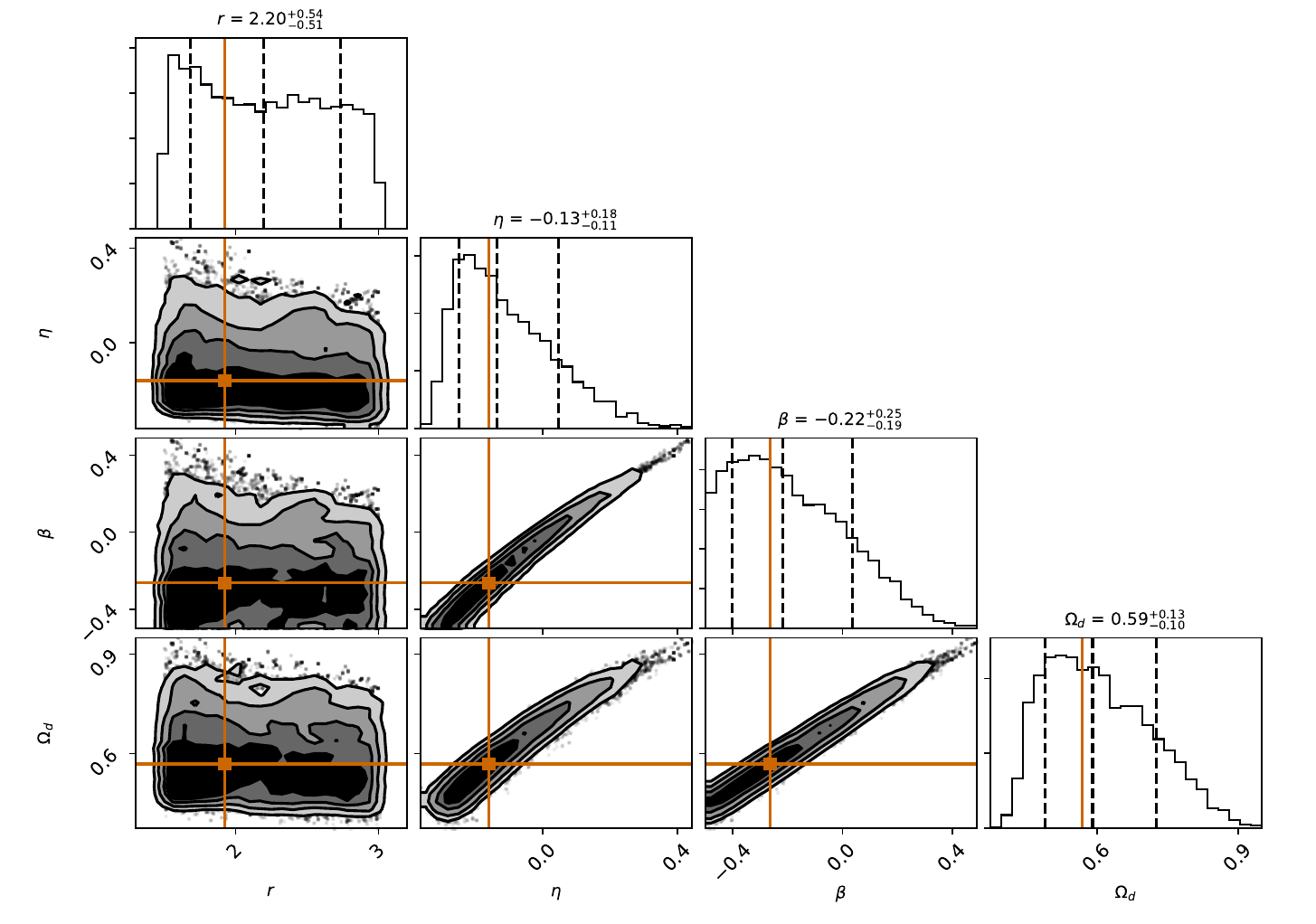} 
\caption{The above depicts the confidence contour for function C4 and model M1.}
\end{center}
\end{figure}

\begin{figure}[htb]
\begin{center}
 \includegraphics[scale=0.5]{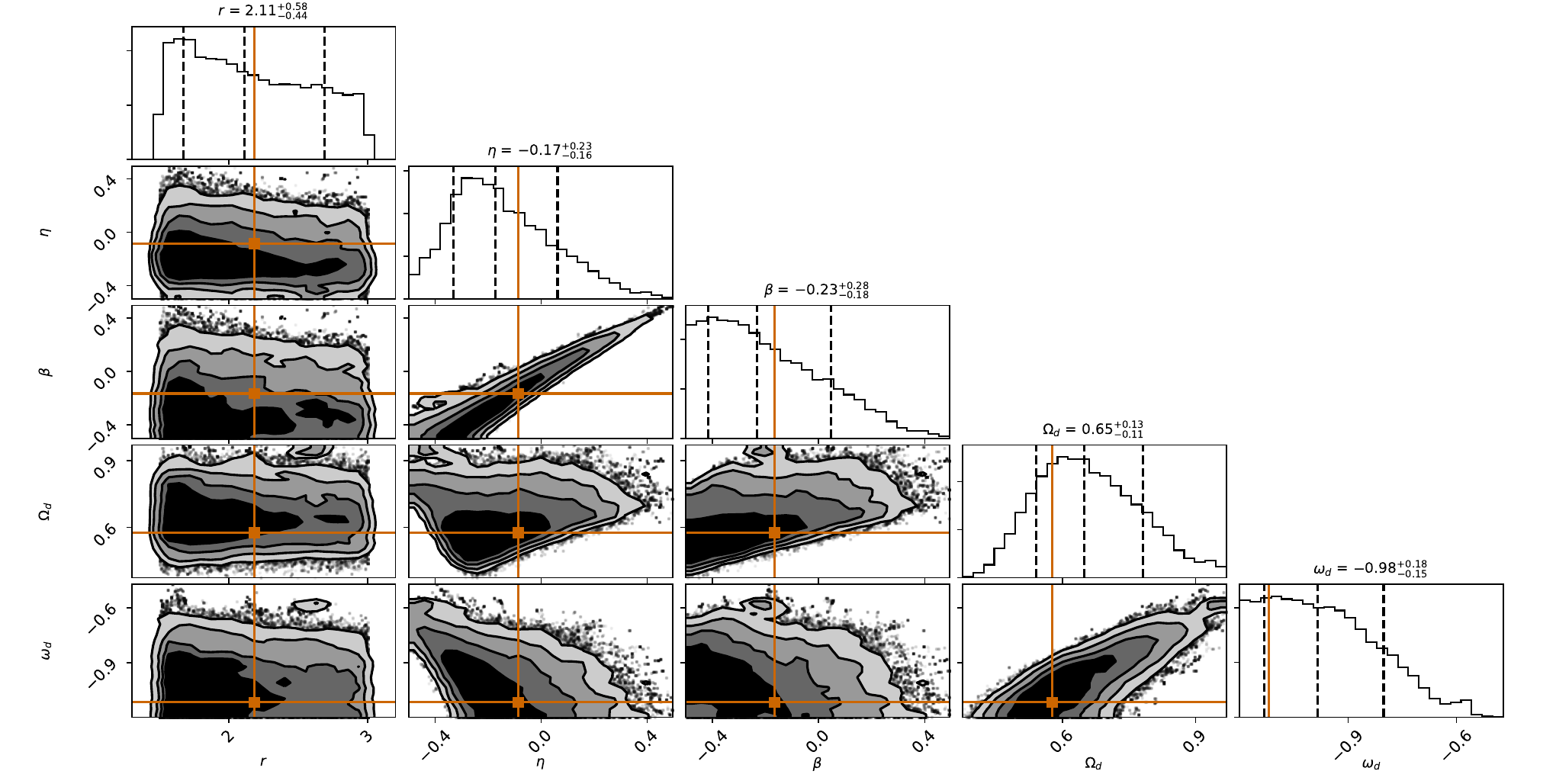}
\caption{The above depicts the confidence contour for function C4 and model M2.}
\end{center}
\end{figure}

\begin{figure}[htb]
\begin{center}
\includegraphics[scale=0.5]{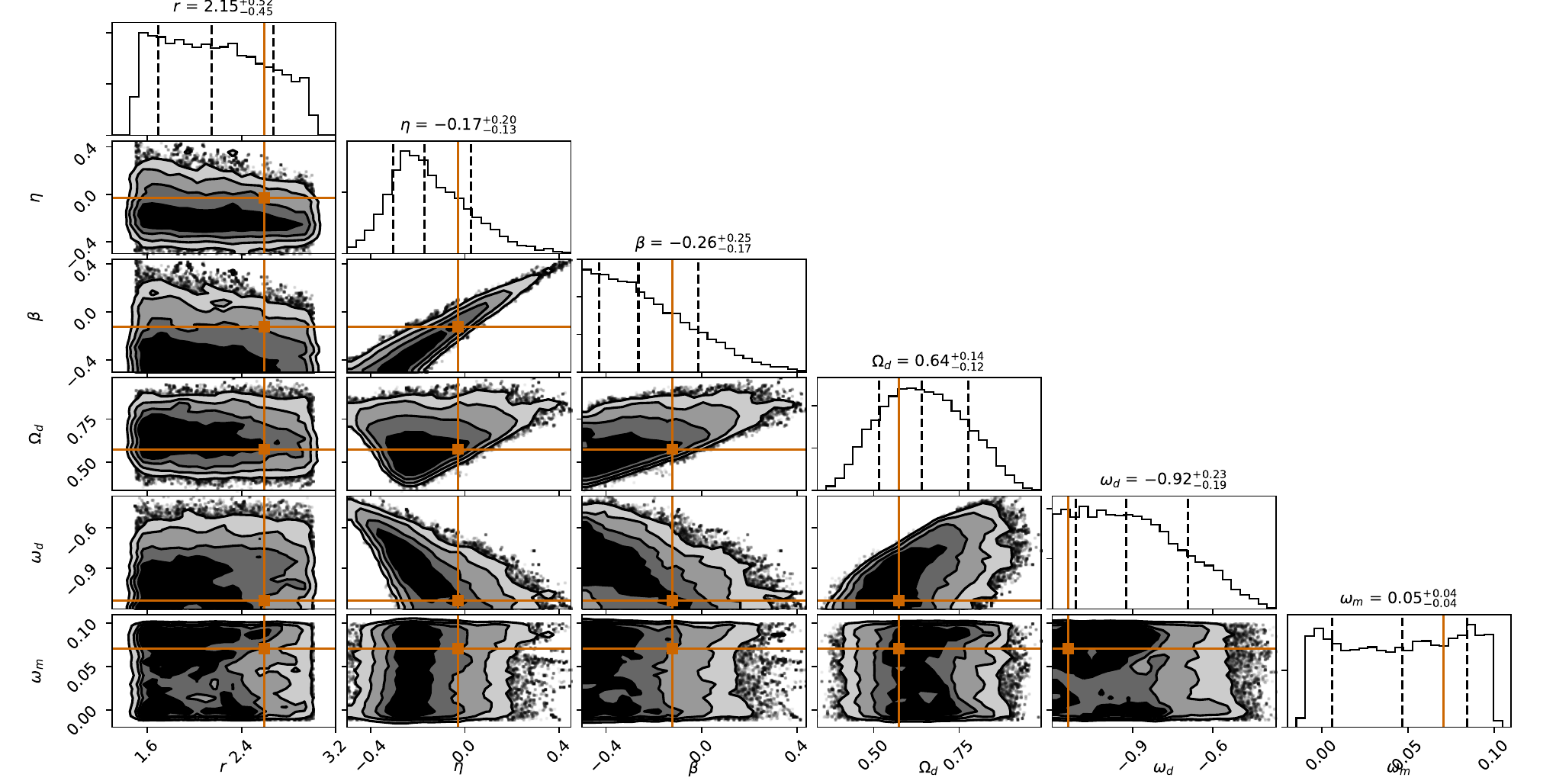}
\caption{The above depicts the confidence contour for function C4 and model M3.}
\end{center}
\end{figure}

\begin{figure}[htb]
\begin{center}
\includegraphics[scale=0.5]{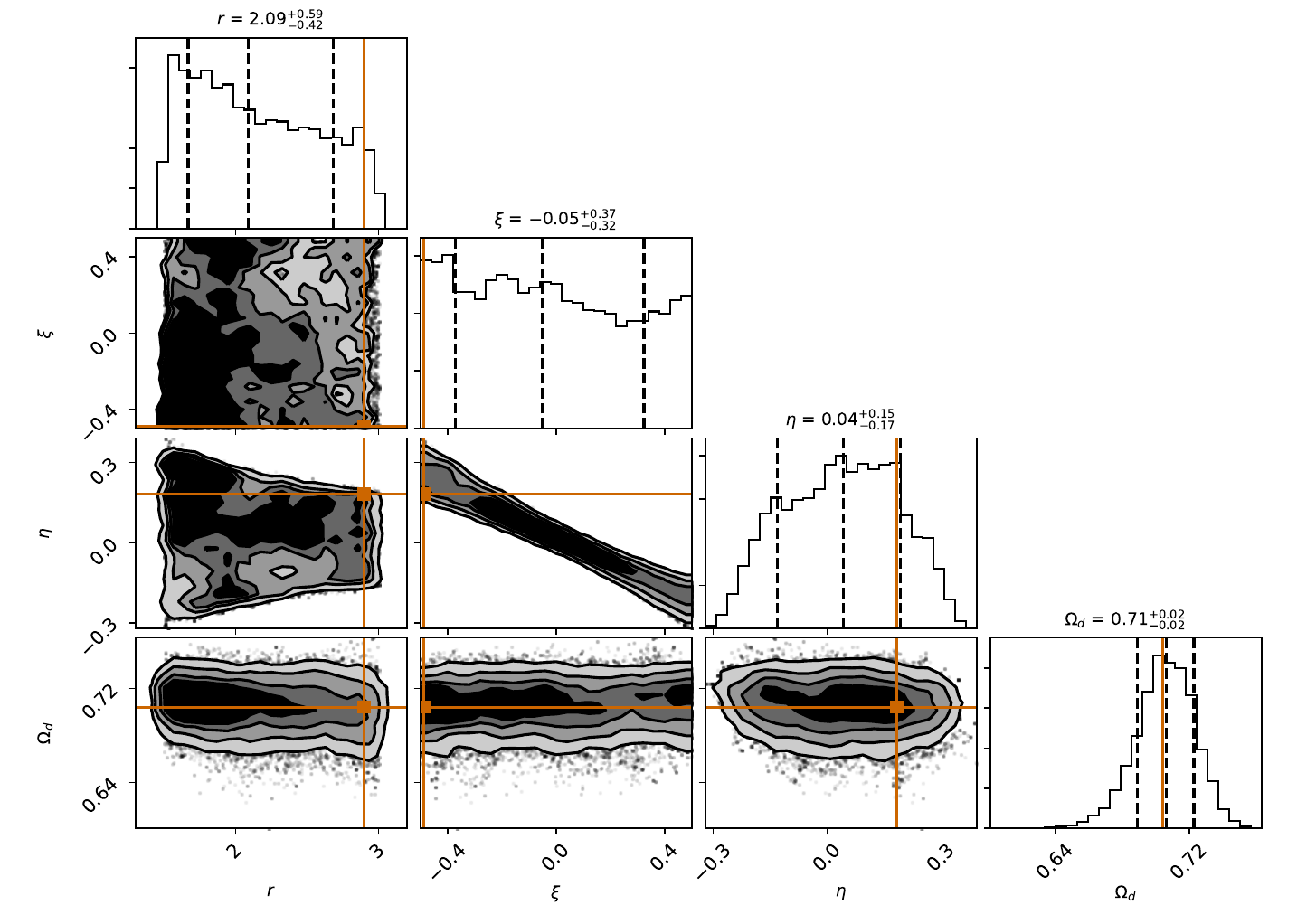}
\caption{The above depicts the confidence contour for function C5 and model M1.}
\end{center}
\end{figure}

\begin{figure}[htb]
\begin{center}
 \includegraphics[scale=0.5]{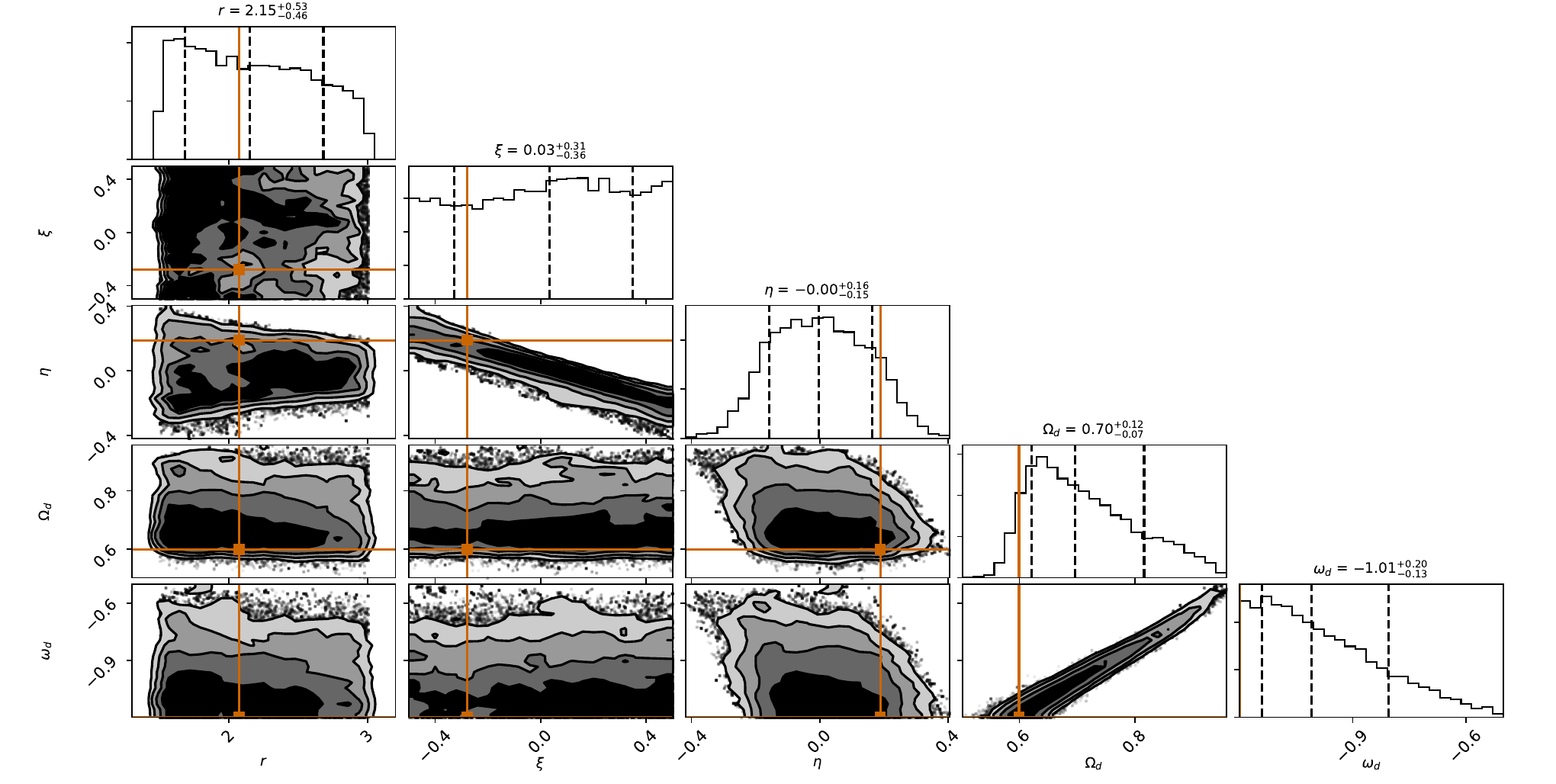}
\caption{The above depicts the confidence contour for function C5 and model M2.}
\end{center}
\end{figure}

\begin{figure}[htb]
\begin{center}
\includegraphics[scale=0.5]{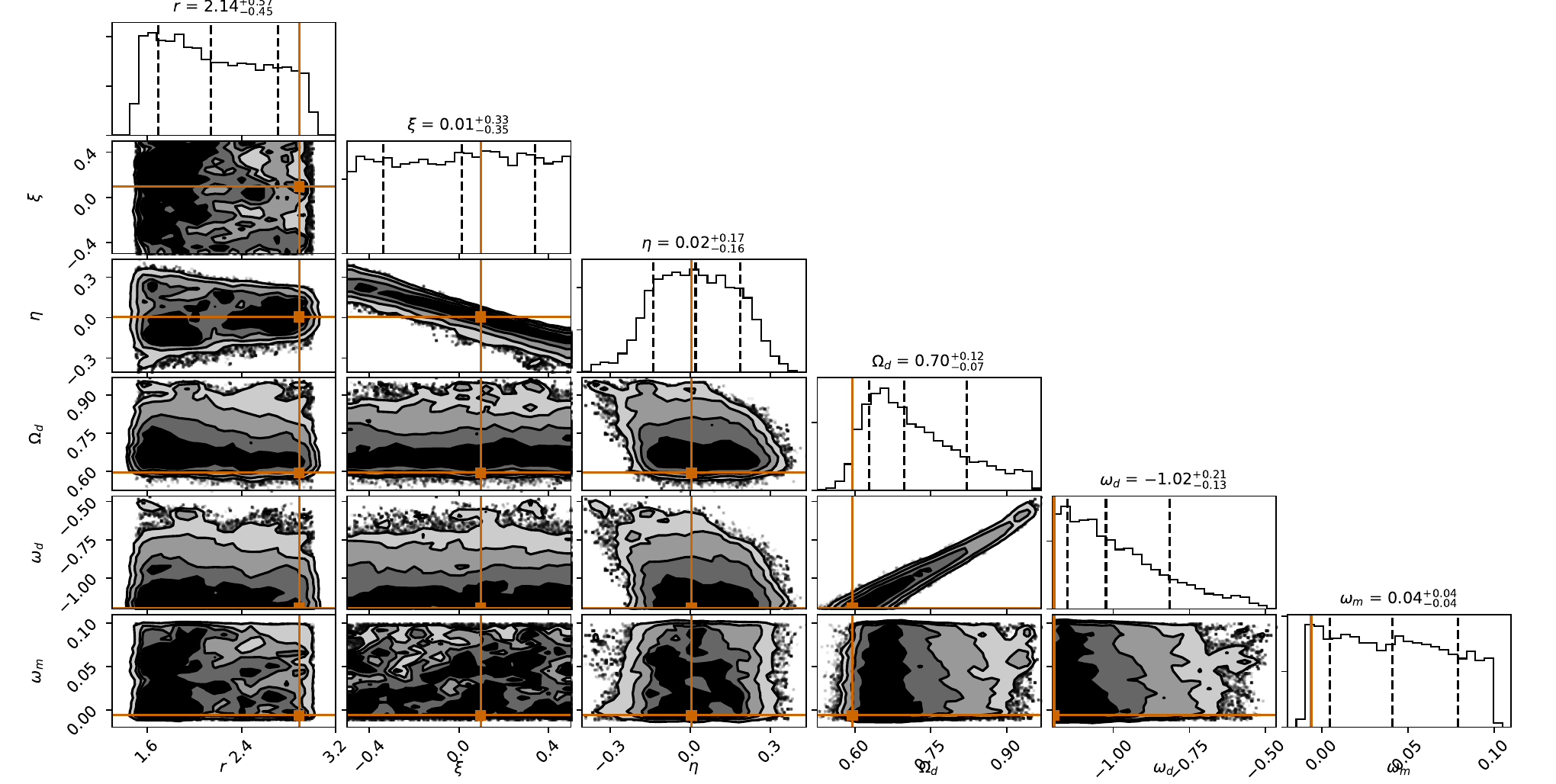}
\caption{The above depicts the confidence contour for function C5 and model M3.}
\end{center}
\end{figure}

\begin{figure}[htb]
\begin{center}
\includegraphics[scale=0.5]{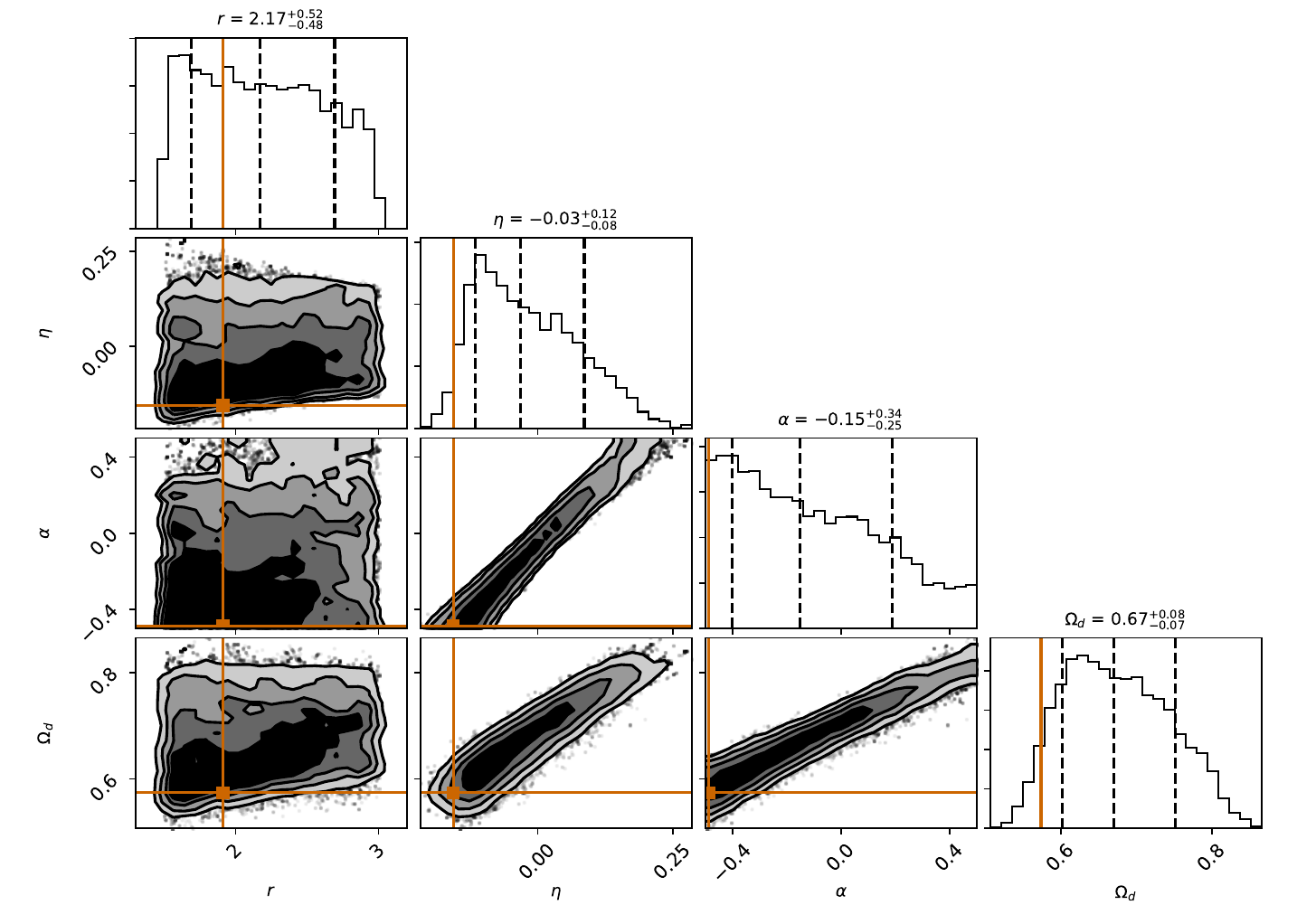}\quad 
\caption{The above depicts the confidence contour for function C6 and model M1.}
\end{center}
\end{figure}

\begin{figure}[htb]
\begin{center}
 \includegraphics[scale=0.5]{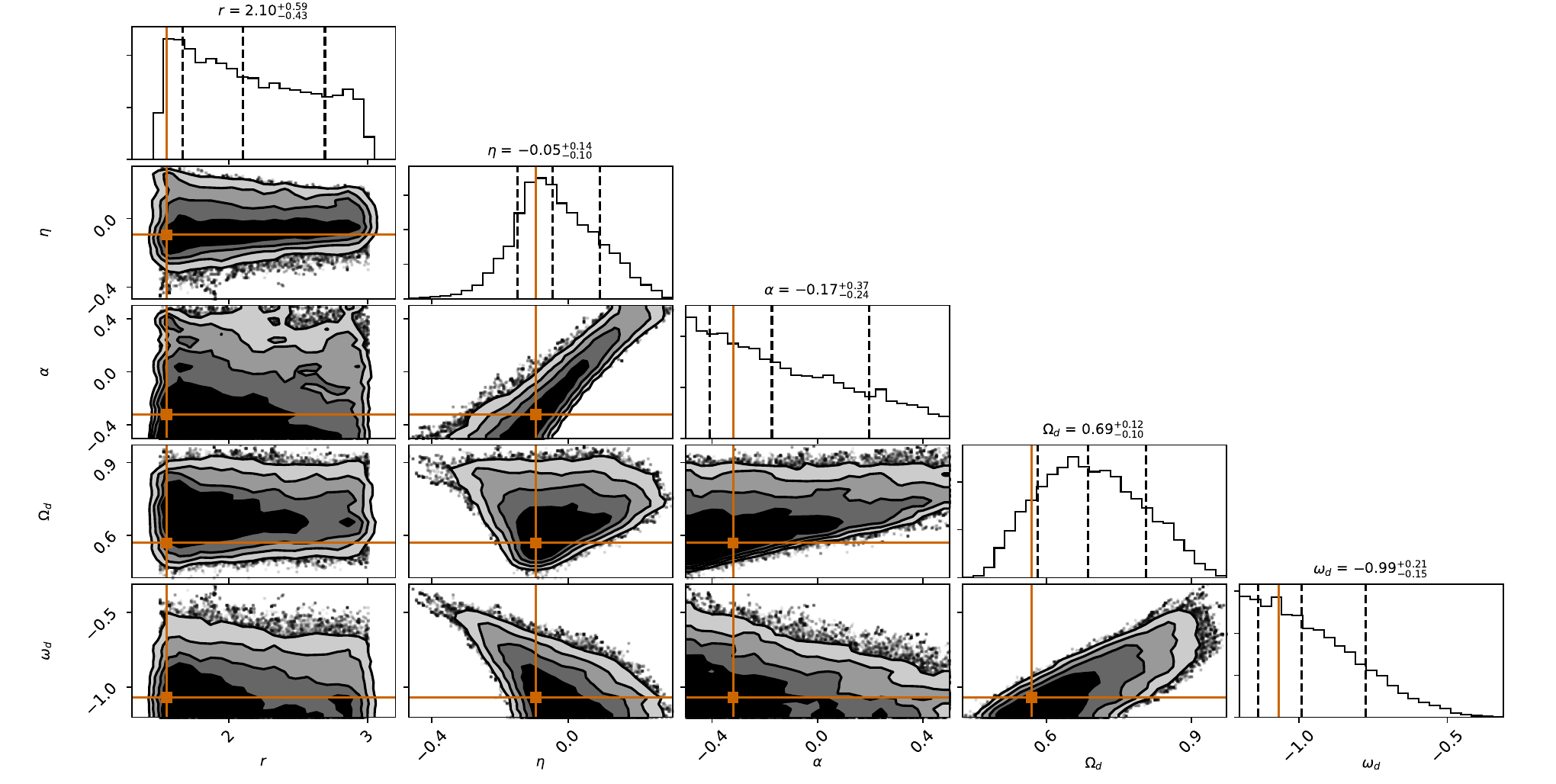}
\caption{The above depicts the confidence contour for function C6 and model M2.}
\end{center}
\end{figure}

\begin{figure}[htb]
\begin{center}
\includegraphics[scale=0.5]{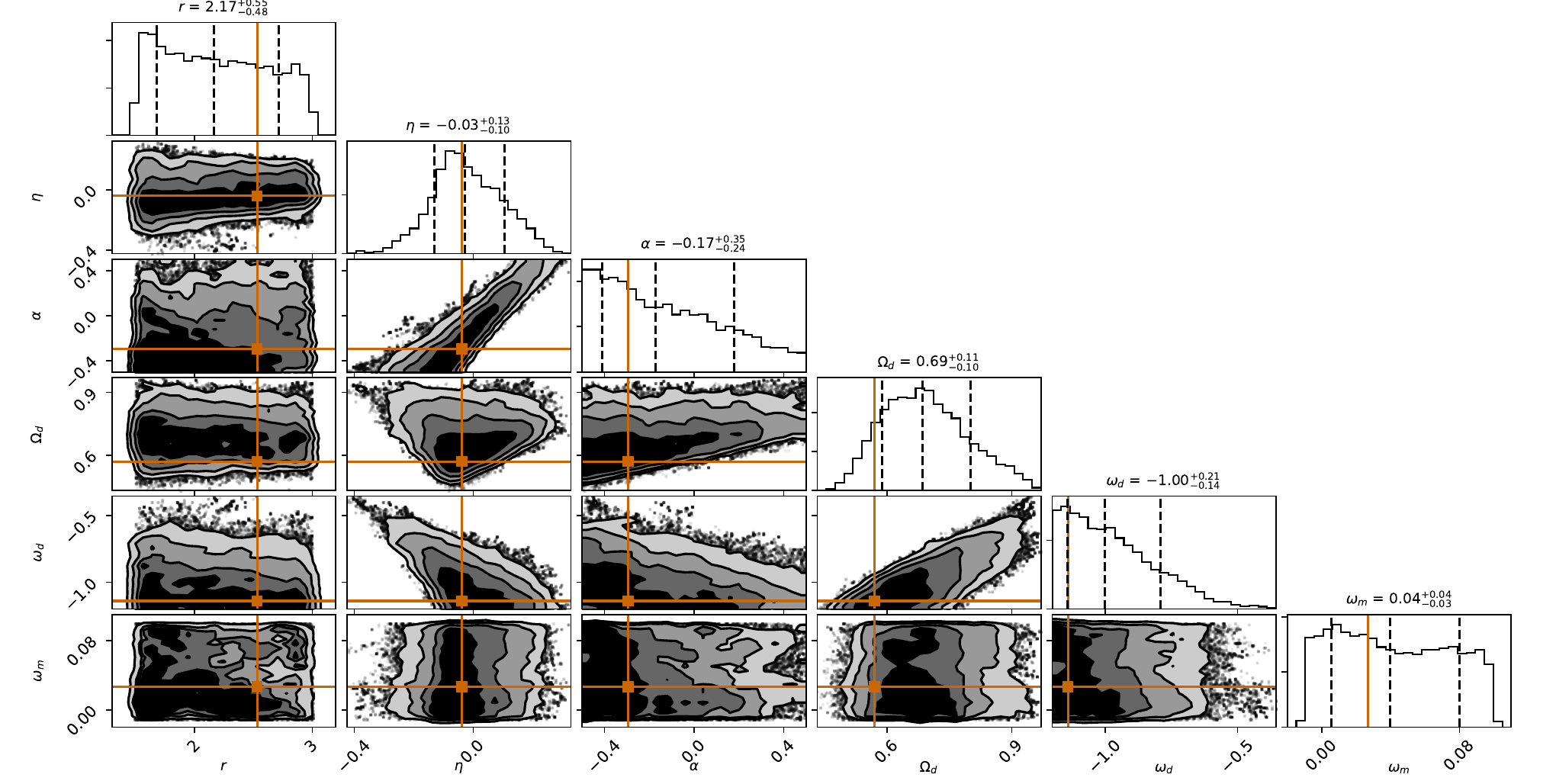}
\caption{The above depicts the confidence contour for function C6 and model M3.}
\end{center}
\end{figure}

\begin{figure}[htb]
\begin{center}
\includegraphics[scale=0.5]{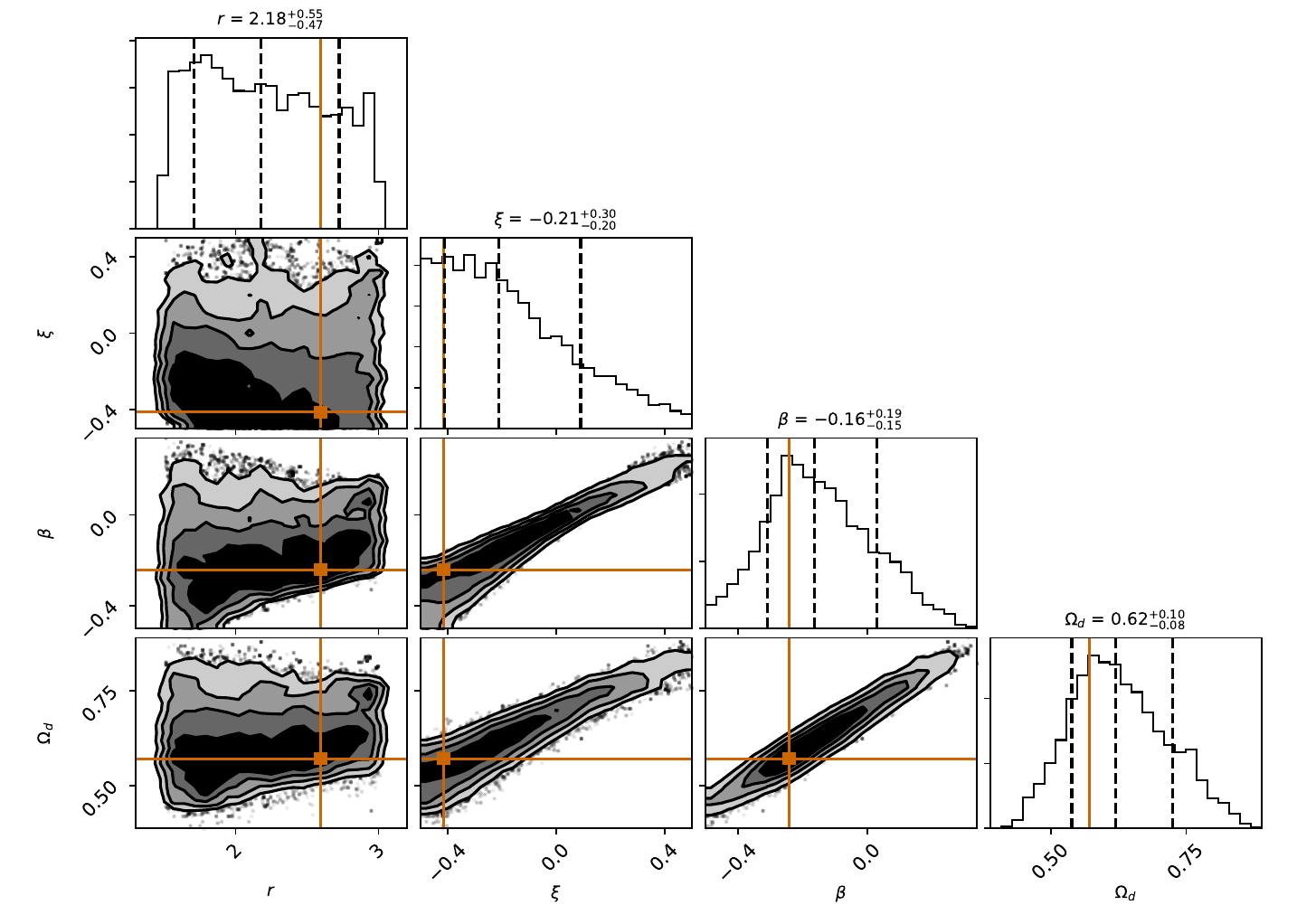}
\caption{The above depicts the confidence contour for function C7 and model M1.}
\end{center}
\end{figure}

\begin{figure}[htb]
\begin{center}
\includegraphics[scale=0.5]{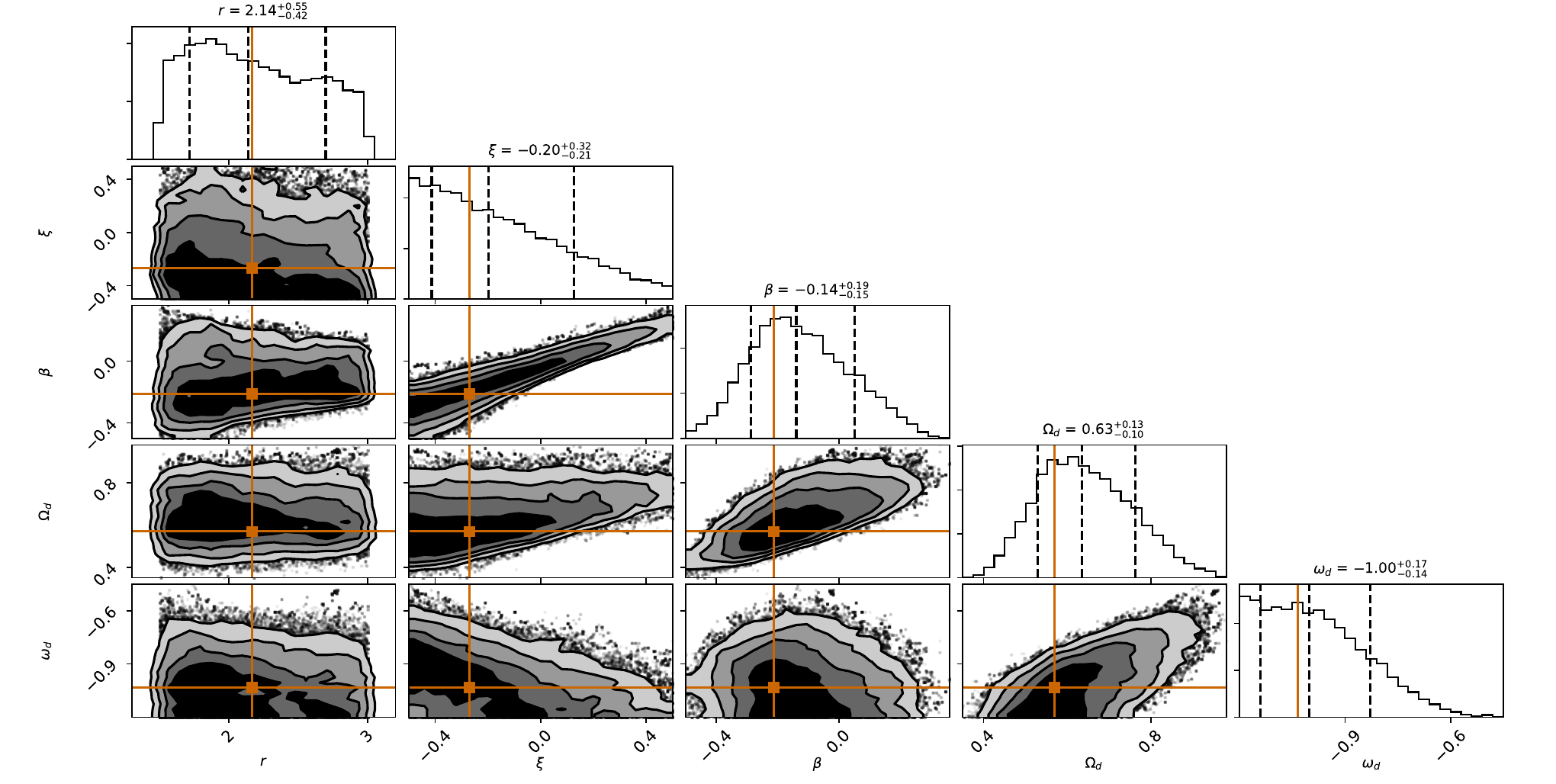}
\caption{The above depicts the confidence contour for function C7 and model M2.}
\end{center}
\end{figure}

\begin{figure}[htb]
\begin{center}
\includegraphics[scale=0.5]{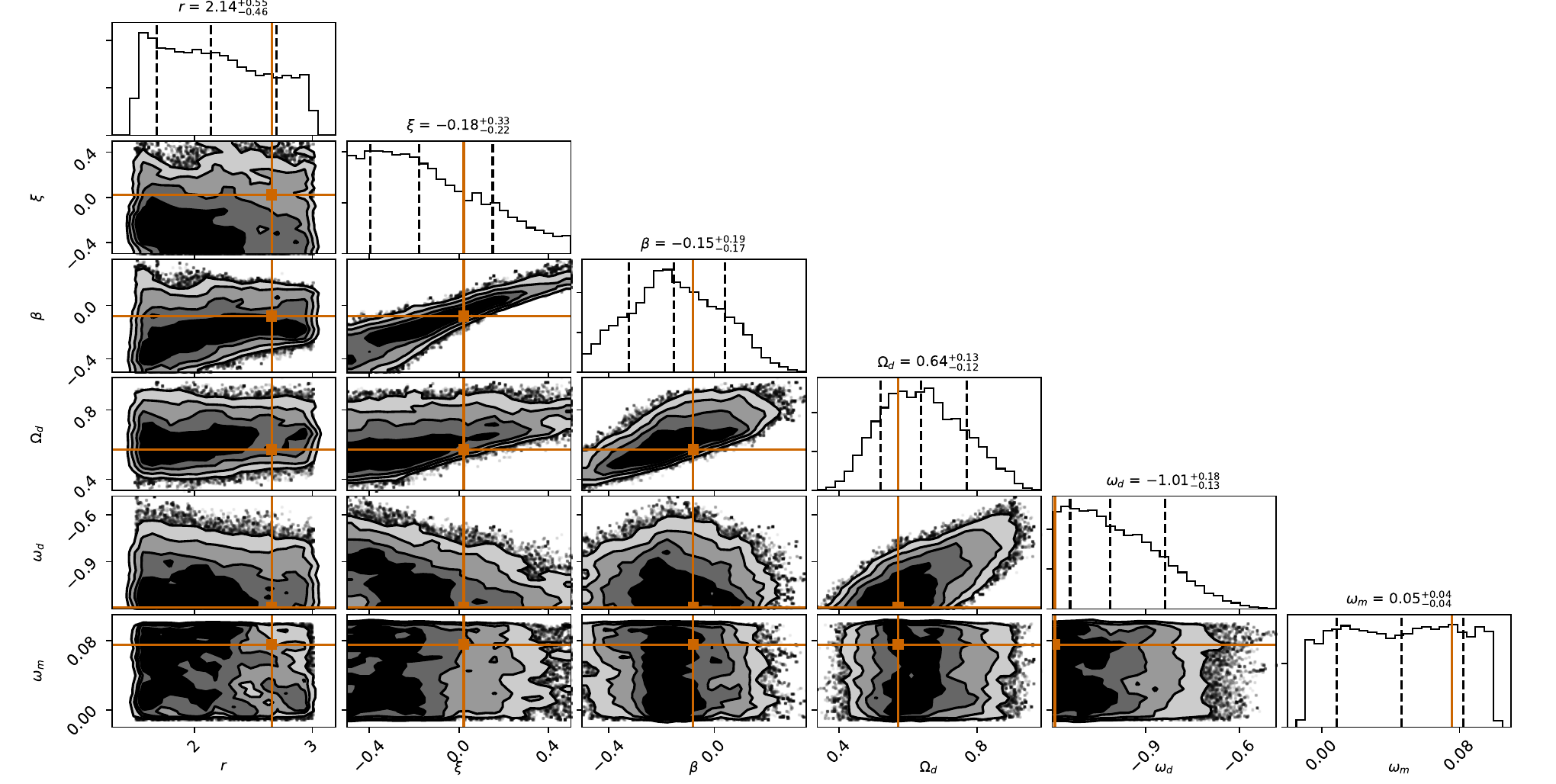}
\caption{The above depicts the confidence contour for function C7 and model M3.}
\end{center}
\end{figure}

\begin{figure}[htb]
\begin{center}
\includegraphics[scale=0.5]{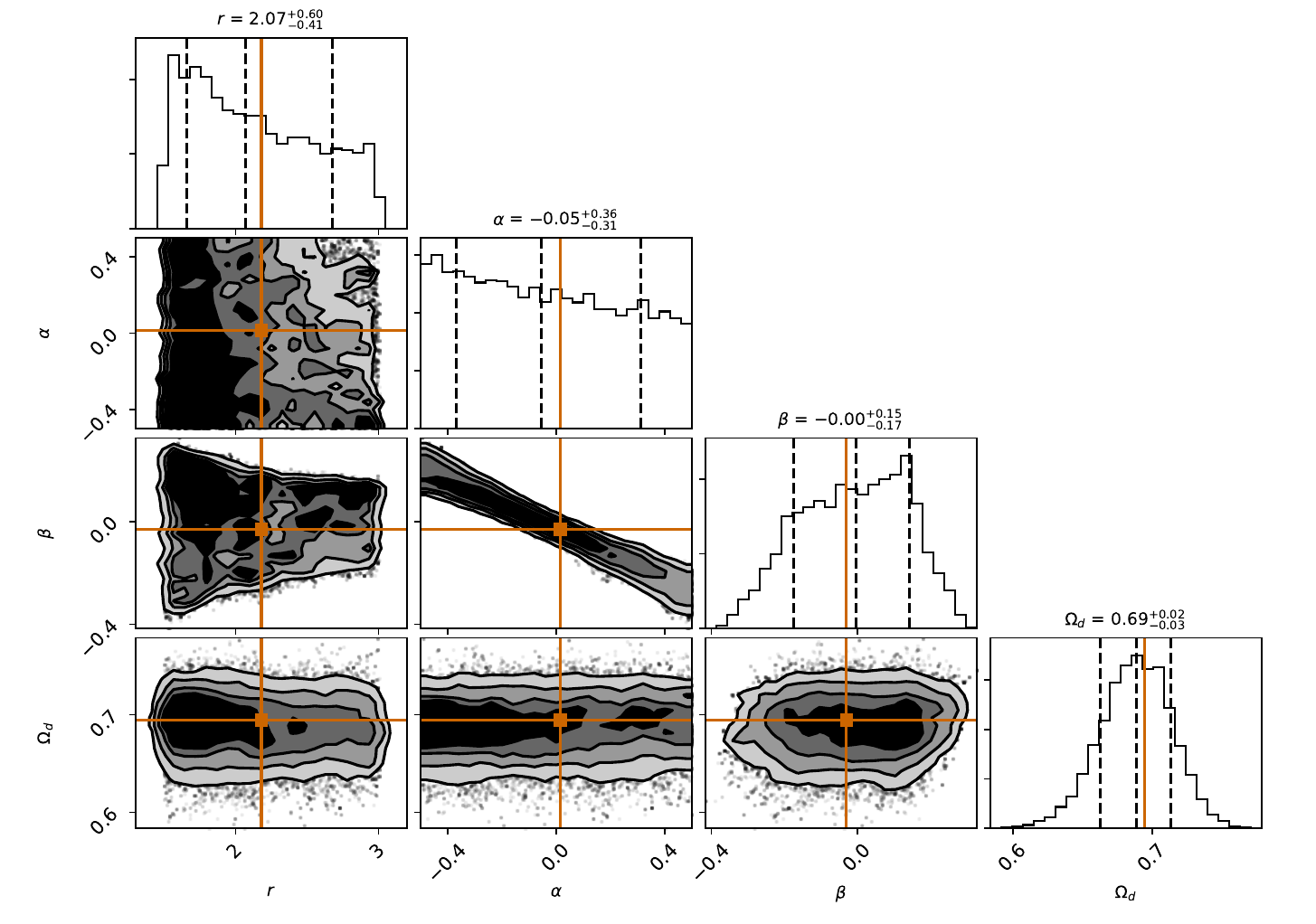}\quad 
\caption{The above depicts the confidence contour for function C8 and model M1.}
\end{center}
\end{figure}

\begin{figure}[htb]
\begin{center}
 \includegraphics[scale=0.5]{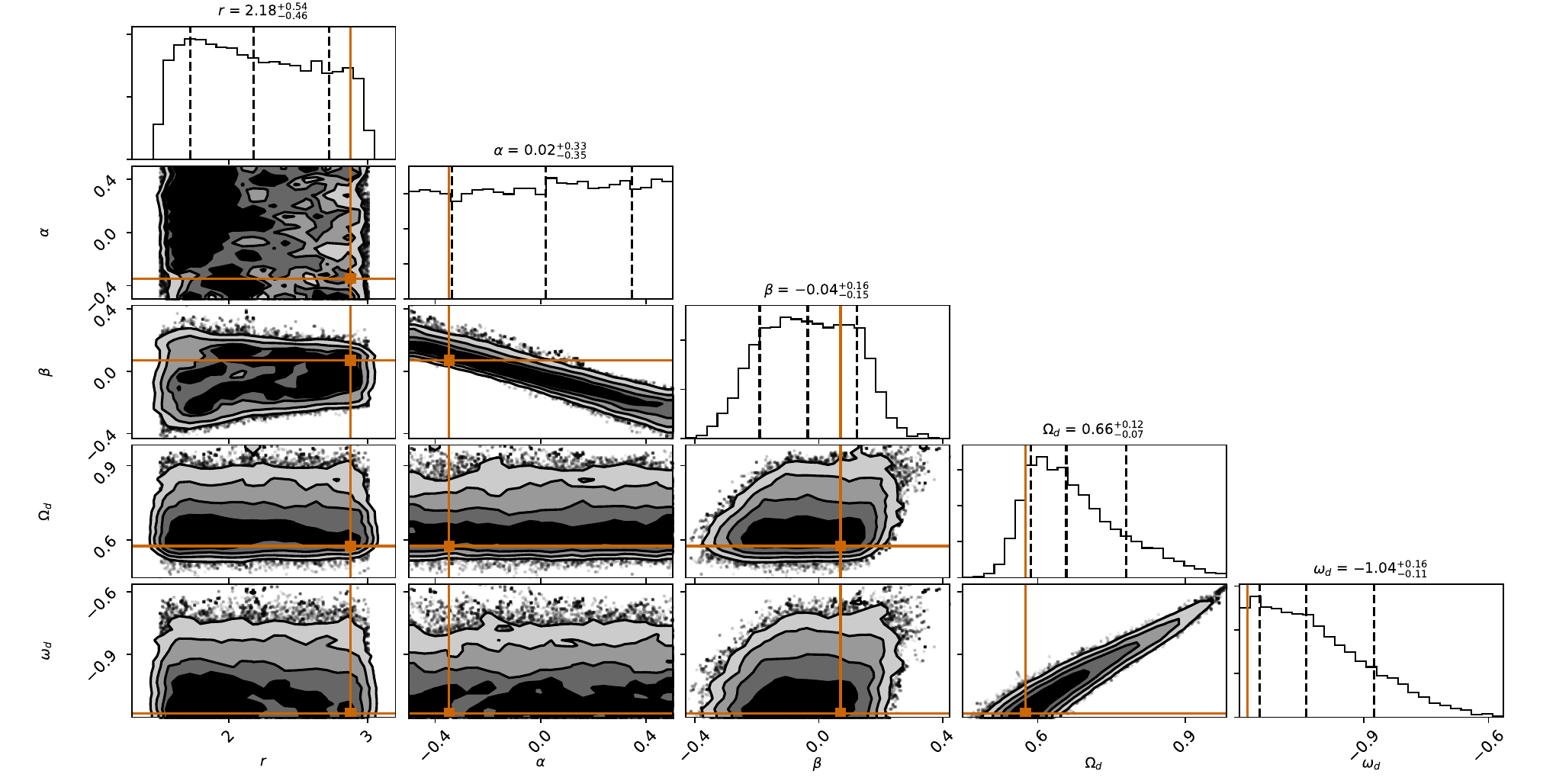}
\caption{The confidence contour for function C8 and model M2.}
\end{center}
\end{figure}

\begin{figure}[htb]
\begin{center}
\includegraphics[scale=0.5]{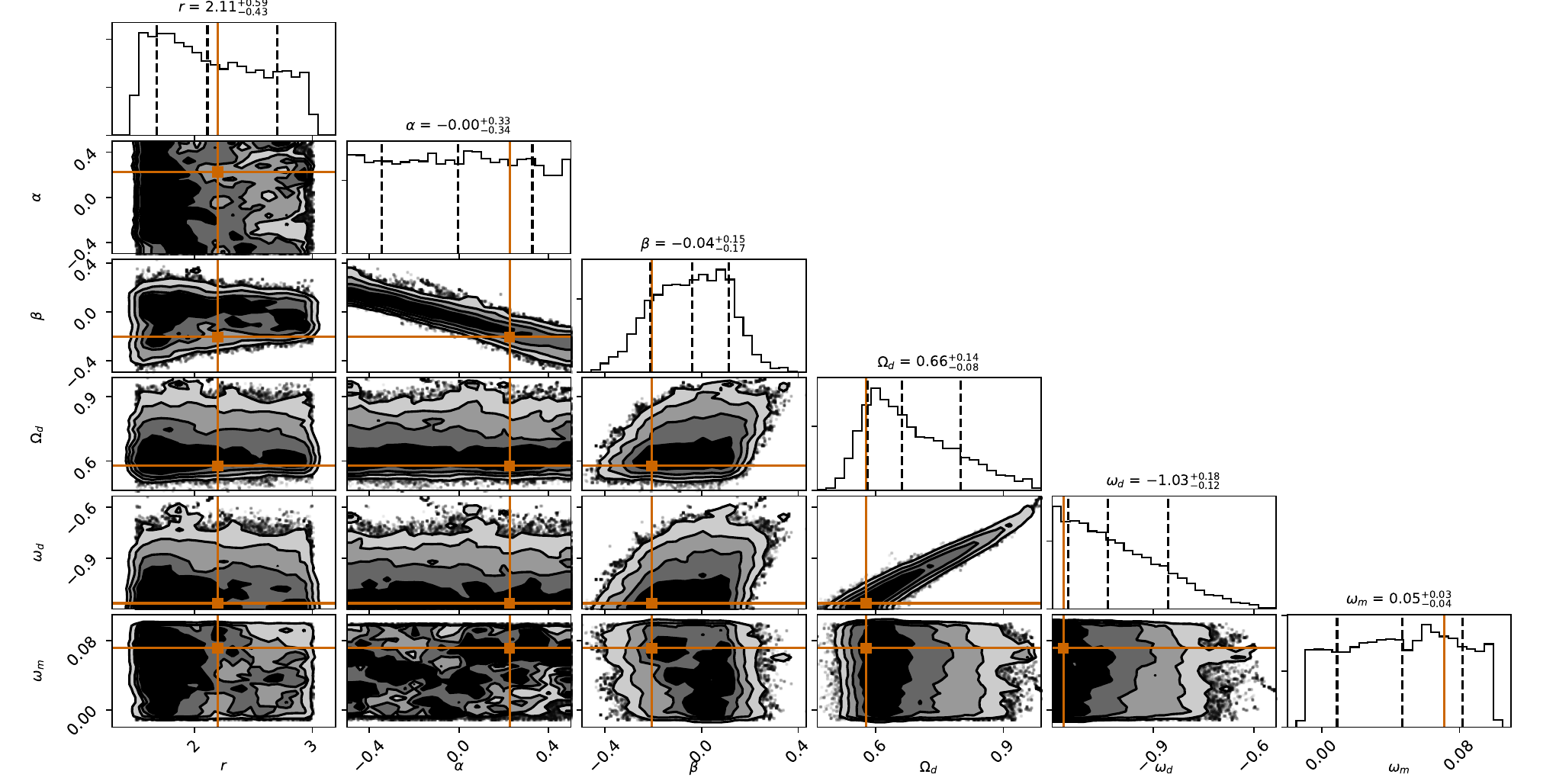}
\caption{The confidence contour for function C8 and model M3.}
\end{center}
\end{figure}

\begin{figure}[htb]
\begin{center}
\includegraphics[scale=0.5]{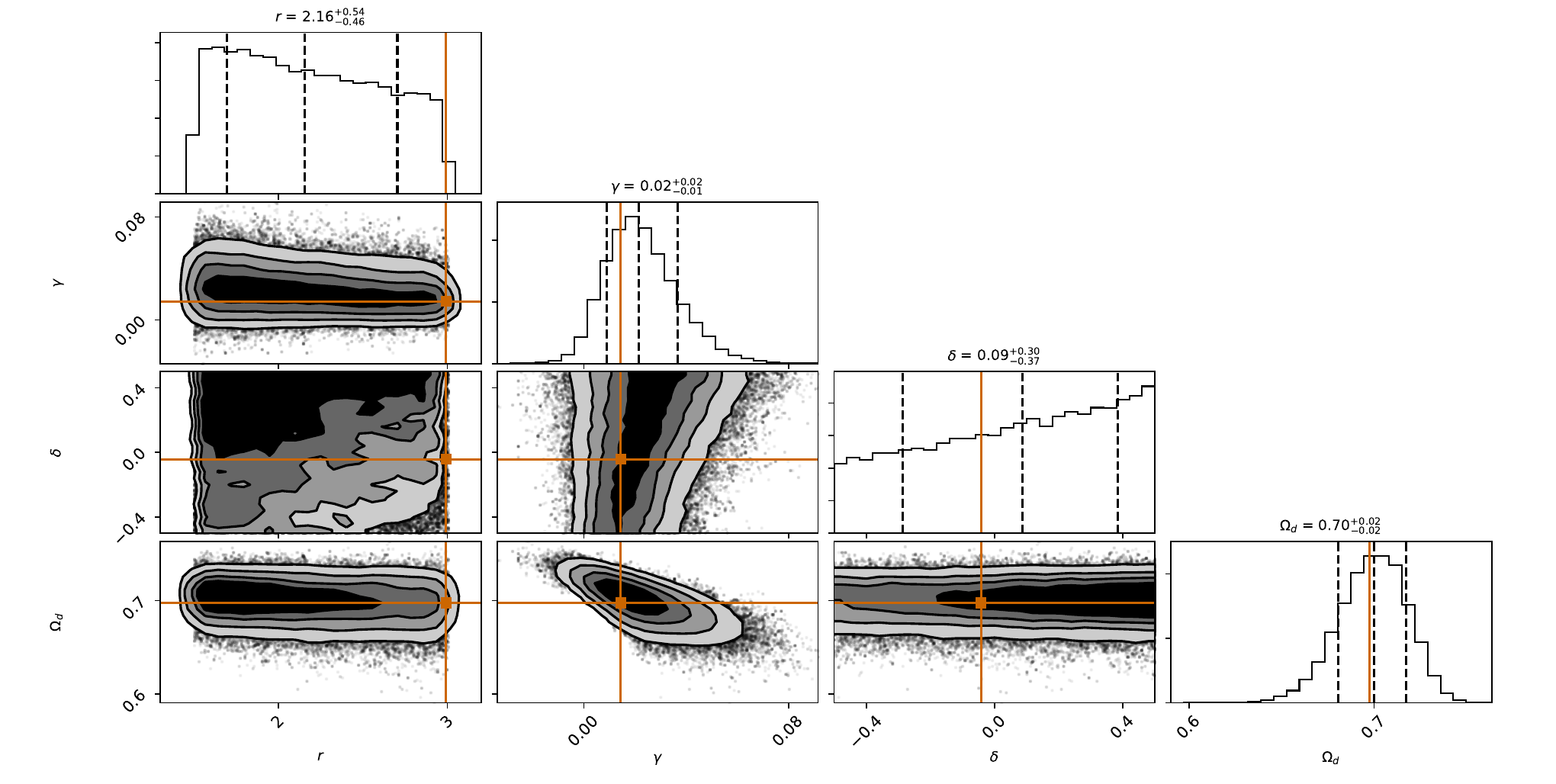}
\caption{The confidence contour for function C9 and model M1.}
\end{center}
\end{figure}

\begin{figure}[htb]
\begin{center}
\includegraphics[scale=0.5]{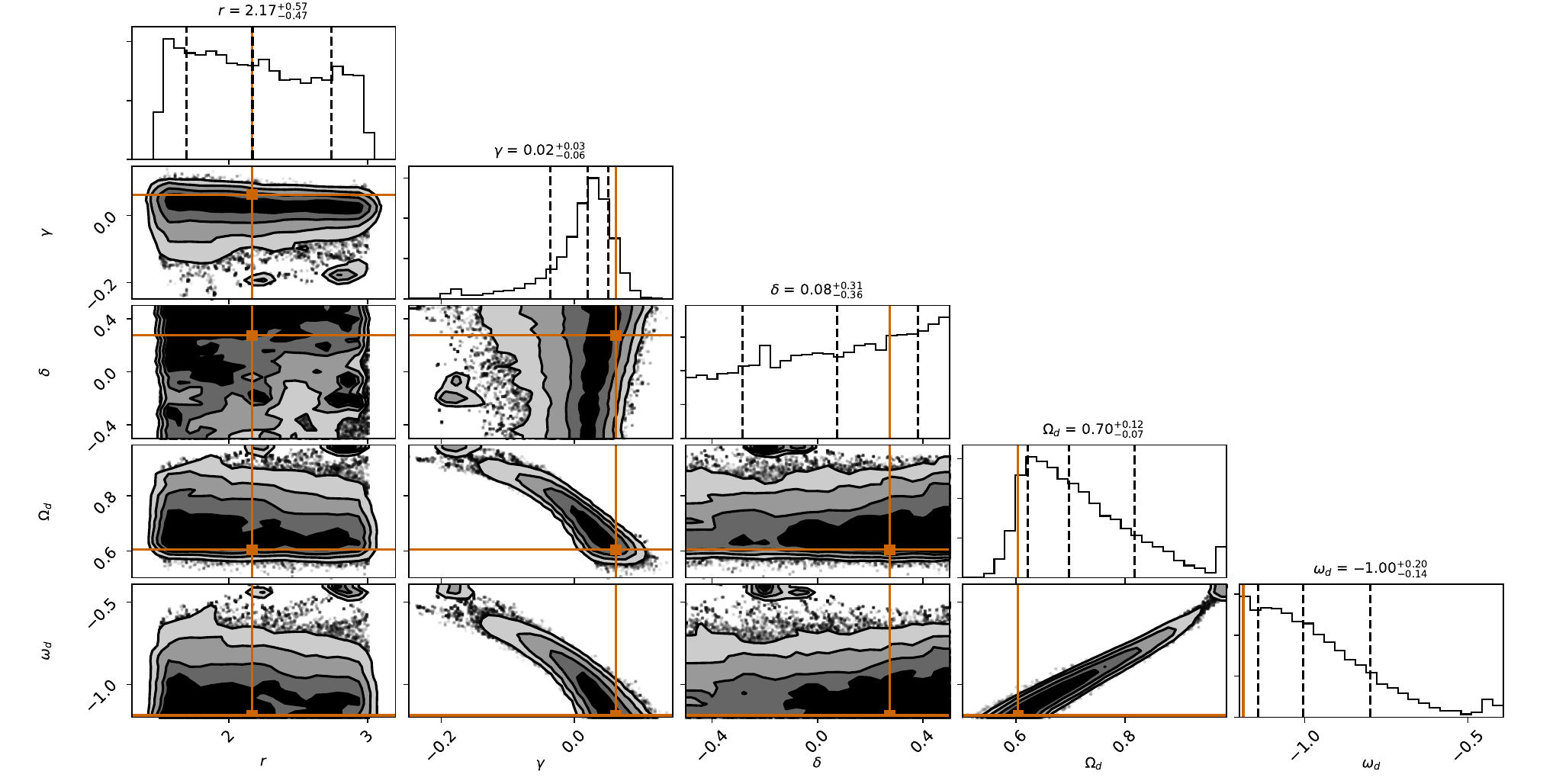}\quad
\caption{The confidence contour for function C9 and model M2.}
\end{center}
\end{figure}

\begin{figure}[htb]
\begin{center}
\includegraphics[scale=0.5]{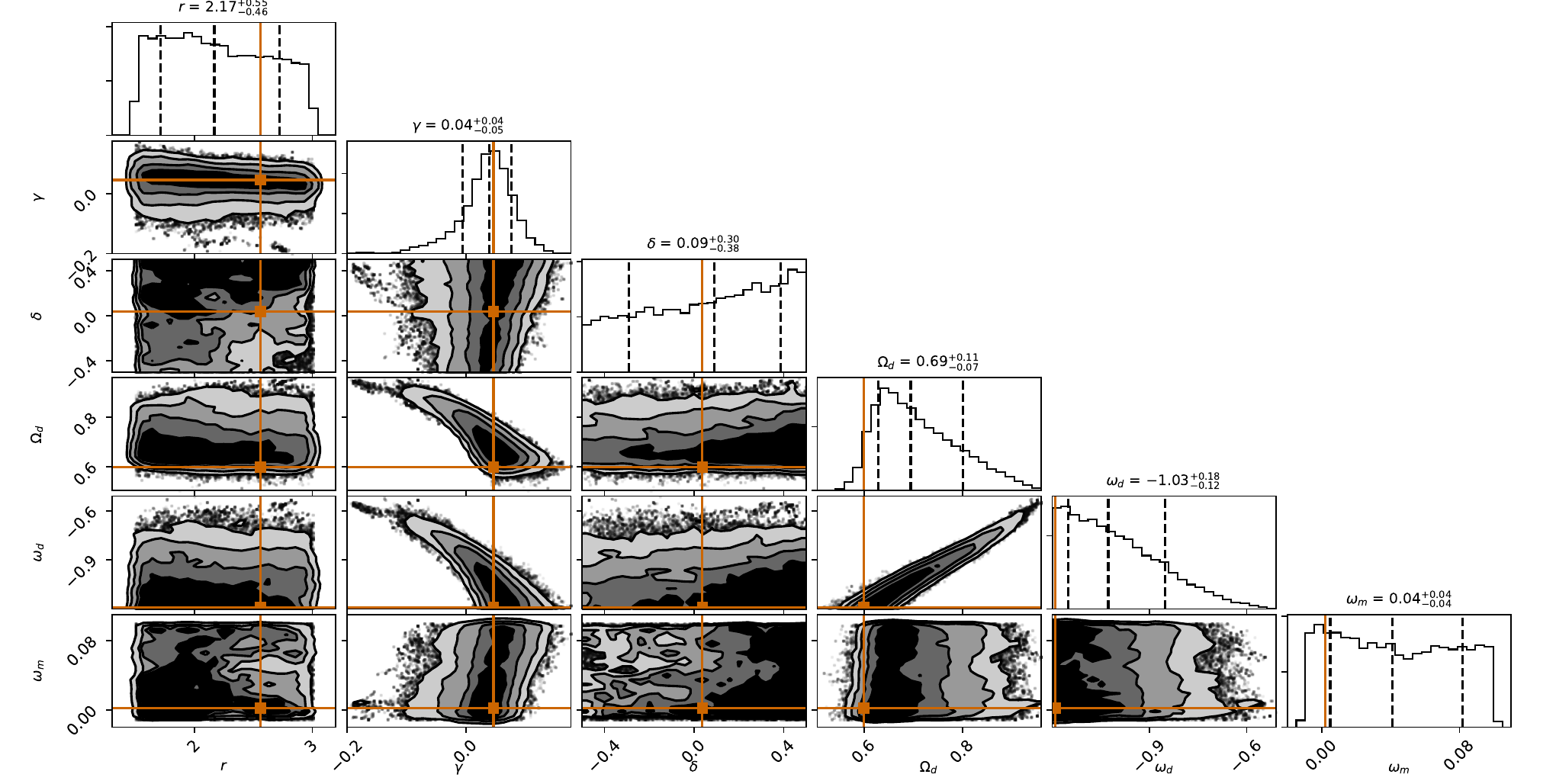}
\caption{The confidence contour for function C9 and model M3.}
\end{center}
\end{figure}

\end{document}